\documentclass[10pt,twocolumn,letterpaper]{article}
\usepackage[final]{cvpr}

\usepackage{graphicx}
\usepackage{amsmath}
\usepackage{amssymb}
\usepackage{booktabs}

\usepackage{multirow}
\usepackage{makecell}
\usepackage{adjustbox}

\usepackage[pagebackref,breaklinks,colorlinks]{hyperref}
\usepackage[capitalize]{cleveref}
\crefname{section}{Sec.}{Secs.}
\Crefname{section}{Section}{Sections}
\Crefname{table}{Table}{Tables}
\crefname{table}{Tab.}{Tabs.}
\def\cvprPaperID{935}
\def\confName{CVPR}
\def\confYear{2023}

\begin{document}

\title{OSRT: Omnidirectional Image Super-Resolution with \\Distortion-aware Transformer\vspace{-4mm}}

\author{Fanghua Yu$^{1*}$ \quad Xintao Wang$^{2}$\thanks{Equal contribution} \quad Mingdeng Cao$^{2, 3}$ \quad Gen Li$^{4}$ \quad Ying Shan$^{2}$ \quad Chao Dong$^{1, 5}$\thanks{Corresponding author (e-mail: chao.dong@siat.ac.cn)}\\
$^{1}$SIAT, Chinese Academy of Sciences \quad $^{2}$ARC, Tencent PCG \quad $^{3}$The University of Tokyo\\
$^{4}$Platform Technologies, Tencent Online Video \quad $^{5}$Shanghai AI Lab\\
{\tt\small fanghuayu96@gmail.com, xintaowang@tencent.com, cmd@g.ecc.u-tokyo.ac.jp}\\
{\tt\small \{genli, yingsshan\}@tencent.com, chao.dong@siat.ac.cn}}

\maketitle

\begin{abstract}
\vspace{-2mm}
Omnidirectional images (ODIs) have obtained lots of research interest for immersive experiences.
Although ODIs require extremely high resolution to capture details of the entire scene, the resolutions of most ODIs are insufficient.
Previous methods attempt to solve this issue by image super-resolution (SR) on equirectangular projection (ERP) images.
However, they omit geometric properties of ERP in the degradation process, and their models can hardly generalize to real ERP images.
In this paper, we propose Fisheye downsampling, which mimics the real-world imaging process and synthesizes more realistic low-resolution samples.
Then we design a distortion-aware Transformer (OSRT) to modulate ERP distortions continuously and self-adaptively.
Without a cumbersome process, OSRT outperforms previous methods by about 0.2dB on PSNR.
Moreover, we propose a convenient data augmentation strategy, which synthesizes pseudo ERP images from plain images.
This simple strategy can alleviate the over-fitting problem of large networks and significantly boost the performance of ODISR.
Extensive experiments have demonstrated the state-of-the-art performance of our OSRT.
\vspace{-2mm}
\end{abstract}

\section{Introduction}
\label{sec:intro}


\begin{figure}[h]
	\small
	\footnotesize
	\vspace{-2mm}
	\begin{tabular}{cc}
				\hspace{-1mm}
				\includegraphics[width=0.22\textwidth]{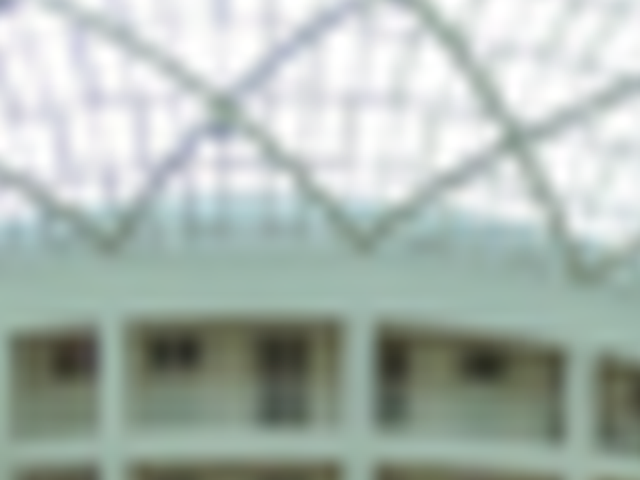} \hspace{-1mm} &
				\includegraphics[width=0.22\textwidth]{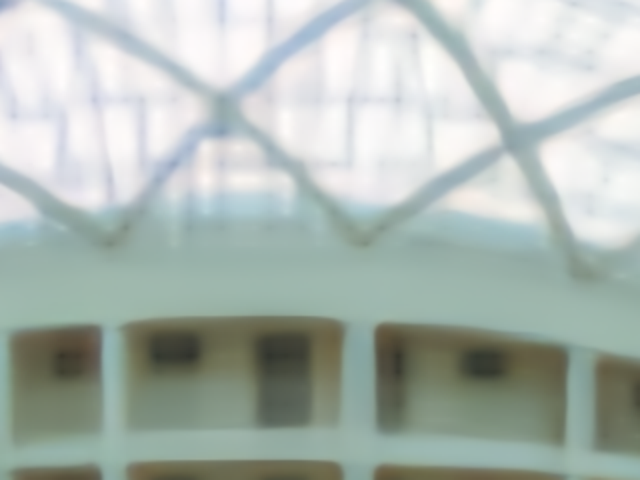} \hspace{-1mm} 
				\\
				\hspace{-1mm}
				Unseen LR \hspace{-1mm} &
				LAU-Net \cite{LauNet} w/o Fisheye  \hspace{-1mm} 
				\vspace{1.5mm}
				\\
				\hspace{-1mm}
				\includegraphics[width=0.22\textwidth]{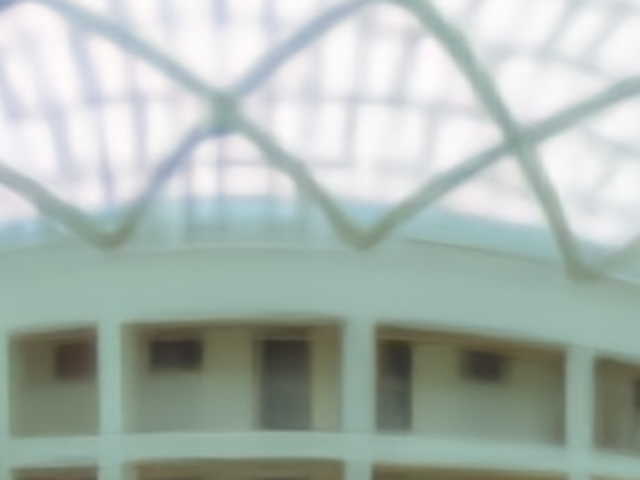} \hspace{-1mm} &
				\includegraphics[width=0.22\textwidth]{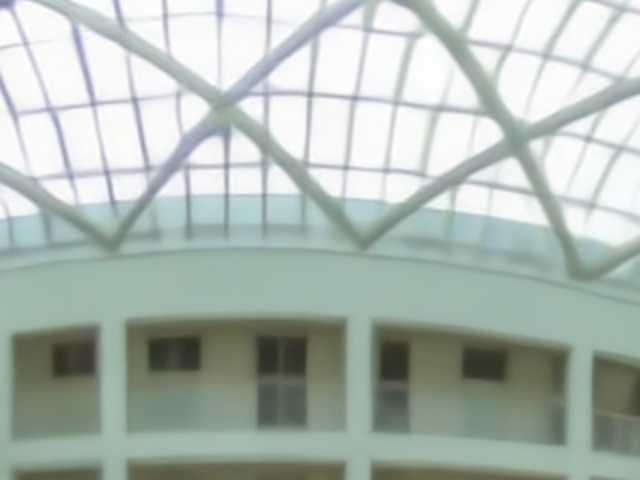} \hspace{-1mm} 
				\\
				\hspace{-1mm}
				OSRT w/o Fisheye \hspace{-1mm} &
				OSRT w/ Fisheye \hspace{-1mm} 
				\\	
	\end{tabular}
	\vspace{-3mm}
	\caption{Visual comparisons of $\times$8 SR results on unseen LR images\protect\footnotemark. Fisheye denotes that the downsampling process in training stages is under Fisheye images.}
	\label{fig:intro_x8_visual}
	\vspace{-8mm}
\end{figure}


In pursuit of the realistic visual experience, omnidirectional images (ODIs), also known as 360$^\circ$ images or panoramic images, have obtained lots of research interest in the computer vision community.
%
%
In reality, we usually view ODIs with a narrow field-of-view (FOV), \textit{e.g.}, viewing in a headset.
%
To capture details of the entire scene, ODIs require extremely high resolution, \textit{e.g.}, 4K $\times$ 8K \cite{ODI_survey}.
However, due to the high industrial cost of camera sensors with high precision, the resolutions of most ODIs are insufficient.
%

Recently, some attempts have been made to solve this problem by image super-resolution (SR) \cite{SISRPre, DistortedPers, Sun360, LauNet, SphereSR}.
As most of the ODIs are stored and transmitted in the equirectangular projection (ERP) type, the SR process is usually performed on the ERP images.
To generate high-/low-resolution training pairs, existing ODISR methods \cite{SISRPre, DistortedPers, Sun360, LauNet, SphereSR} directly apply uniform bicubic downsampling on the original ERP images (called ERP downsampling), which is identical to general image SR settings \cite{EDSR,RCAN}.
While omitting geometric properties of ERP in the degradation process, their models can hardly generalize to real ERP images. 
We can observe missing structures and blur textures in \cref{fig:intro_x8_visual}.\footnotetext{Photoed by Peter Leth on Flickr, with \href{https://creativecommons.org/licenses/by/2.0/}{CC license}.}
%
%
Therefore, we need a more appropriate degradation model before studying SR algorithms.
In practice, ODIs are acquired by the fisheye lens and stored in ERP.
Given that the low-resolution issue in real-world scenarios is caused by insufficient sensor precision and density, the downsampling process should be applied to original-formatted images before converting into other storage types.
Thus, to be conformed with real-world imaging processes, we propose to apply uniform bicubic downsampling on Fisheye images, which are the original format of ODIs.
The new downsampling process (called Fisheye downsampling) applies uniform bicubic downsampling on Fisheye images before converting them to ERP images.
Our Fisheye downsampling is more conducive to exploring the geometric property of ODIs.


The key issue of ODISR algorithm design is to utilize the geometric properties of ERP images, which is also the focus of previous methods.
For example, Nishiyama \textit{et al.} \cite{DistortedPers} add a distortion-related condition as an additional input.
LAU-Net \cite{LauNet} splits the whole ERP image into patches by latitude band and learns upscaling processes separately.
However, the separated learning process will lead to information disconnection between adjacent patches.
SphereSR \cite{SphereSR} learns different upscaling functions on various projection types, but will inevitably introduce multiple-time computation costs.
%
%
To push the performance upper bound, we propose the first Transformer for Omnidirectional image Super-Resolution (OSRT), and incorporate geometric properties in a distortion-aware manner.
Specifically, to modulate distorted feature maps, we implement feature-level warping, in which offsets are learned from latitude conditions.
In OSRT, we introduce two dedicated blocks to adapt latitude-related distortion: distortion-aware attention block (DAAB), and distortion-aware convolution block (DACB).
DAAB and DACB are designed to perform distortion modulation in arbitrary Transformers and ConvNets.
These two blocks can directly replace the multi-head self-attention block and convolution layer, respectively.
The benefit of DAAB and DACB can be further improved when being inserted into the same backbone network. 
OSRT outperforms previous methods by about 0.2dB on PSNR (\cref{tab:erp_setting}).


However, the increase of network capacity will also enlarge the overfitting problem of ODISR, which is rarely mentioned before.
The largest ODIs dataset \cite{LauNet} contains only 1K images, which cannot provide enough diversity for training Transformers.
Given that acquiring ODIs requires expensive equipment and tedious work, we propose to generate distorted ERP samples from plain images for data augmentation.
In practice, we regard a plain image as a sampled perspective, and project it back to the ERP format.
Then we can introduce $146$K additional training patches, 6 times of the previous dataset. 
This simple strategy can significantly boost the performance of ODISR (\cref{tab:aug}) and alleviate the over-fitting problem of large networks (\cref{fig:overfit}).
A similar data augmentation method is also applied in Nishiyama \textit{et al.} \cite{DistortedPers}, but shows marginal improvement on small models under ERP downsampling settings. 


Our contributions are threefold.
\textbf{1}) For \textit{problem formulation}: To generate more realistic ERP low-resolution images, we propose Fisheye downsampling, which mimics the real-world imaging process.
\textbf{2}) For \textit{method}: Combined with the geometric properties of ERP, we design a distortion-aware Transformer, which modulates distortions continuously and self-adaptively without cumbersome process.
\textbf{3}) For \textit{data}: To reduce overfitting, we propose a convenient data augmentation strategy, which synthesizes pseudo ERP images from plain images.
Extensive experiments have demonstrated the state-of-the-art performance of our OSRT.

\section{Related Work}
\label{sec:Related Work}


\textbf{Single Image Super-Resolution (SISR).}
Deep learning for single image SR (SISR) is first introduced in \cite{SRCNN}.
Further works boost SR performance by CNNs \cite{FSRCNN,EDSR,RCAN,SAN,HAN,NLSN,BSRN}, Vision Transformers (ViTs) \cite{SwinIR,IPT,EDT,HAT} and generative adversarial networks (GANs) \cite{SRGAN,ESRGAN,RealESRGAN,ranksrgan}.
For instance, EDSR \cite{EDSR} removes Batch Normalization layers and applies a more complicated residual block.
RCAN \cite{RCAN} introduces channel-wise attention mechanisms to a deeper network.
SwinIR \cite{SwinIR} proposes an image restoration Transformer based on \cite{SwinTrans}.
To improve perceptual quality, adversarial training are performed as a tuning process to generate more realistic results \cite{ESRGAN, RealESRGAN}.
Moreover, various flexible degradation models are proposed in \cite{BSRGAN,RealESRGAN} to synthesize more practical degradations.


\textbf{Omnidirectional Image Super-Resolution (ODISR).}
Initially, ODISR models focus on the spherical assembling of LR ODIs under various projection types \cite{previous_odisr_0,previous_odisr_1,previous_odisr_2,previous_odisr_3,previous_odisr_4}.
Recent ODISR models are performed on plane images and are fine-tuned from existing SISR models with L1 loss \cite{SISRPre} or GAN loss \cite{360SS, GAN_Fq}.
The improvements are limited, for they only concern the distribution gap between ODIs and plain images.
Since LAU-Net \cite{LauNet} found pixel density in ERP ODIs is non-uniform, many studies attempt to design specific backbone networks to overcome this issue.
LAU-Net \cite{LauNet} manually splits the whole ERP image into latitude-related patches and learns ERP distortion over different latitude ranges separately.
While LAU-Net learns latitude-related ERP distortion somewhat, its non-overlapped patches lead to disconnection in whole ERP images.
Nishiyama \textit{et al.} \cite{DistortedPers} treats area stretching ratio as additional input.
However, these conditions are tough to be utilized with an unmodified SISR backbone network.
SphereSR \cite{SphereSR} learns upsampling processes on various projection types (CP, ERP, Polyhedron) to mitigate the influence of non-uniformity in specific projection types.
It applies a local implicit image function (LIIF) \cite{LIIF} to query RGB values on spherical surfaces continuously.
Although SphereSR improves information consistency between various ODI projection types, they apply multiple networks to learn the upscaling process of each projection type.
Given that all other projection types in SphereSR are converted from ERP, patterns under various types are reusable when distortions are properly rectified.
Moreover, the complex and unstructured image data in polyhedron projection hinders further research of ODISR.


\textbf{Deformable Mechanism.}
Dai \textit{et al.} \cite{360SS} first propose deformable convolutions to obtain information out of its regular neighborhood.
Xia \textit{et al.} \cite{DeformViT} further verified that Vision Transformers also benefit from applying deformable mechanisms on self-attention blocks.
In Video SR tasks, the deformable mechanism can be adapted to align features between adjacent frames \cite{EDVR,basicvsr,basicvsr++}.

\section{Method}
\label{sec:Method}

In this section, we first analyze the cause of ERP and Fisheye distortions, as well as the relationship between these two distortions (\cref{subs:distortion}).
Then, we discuss the designs of Fisheye downsampling (\cref{subs:fisheye_downsampling}), distortion-aware Transformer (OSRT) (\cref{subs:OSRT}), and the convenient data augmentation strategy (\cref{subs:dataaug}).

\subsection{Revisiting Distortions in ODIs}\label{subs:distortion}

\begin{figure}[ht]
	\vspace{-3mm}
	\centering
	\includegraphics[]{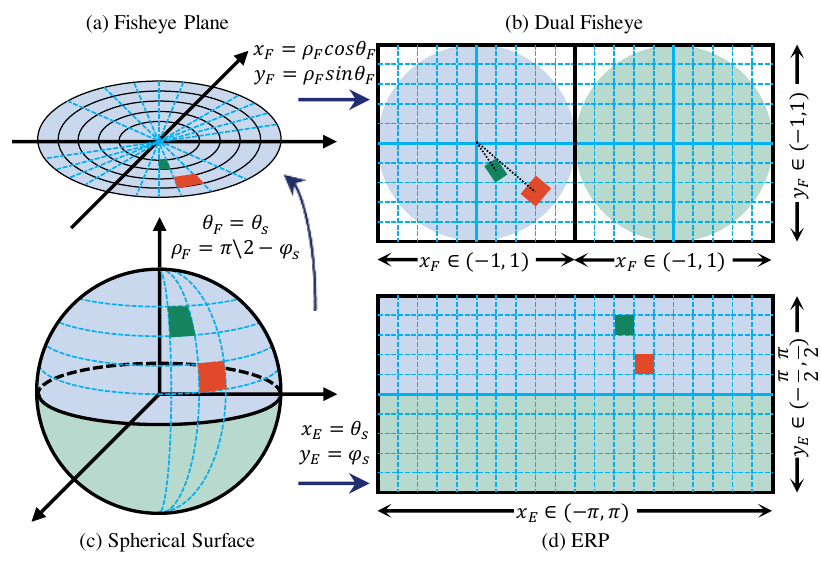}
	\vspace{-7mm}
	\caption{Geometric explanation of transforming between ERP, Fisheye, and the ideal spherical surface. To simplify, we discuss the horizontal spliced Fisheye with an aperture degree of $\pi$.}
	\label{fig:erp_distortion}
	\vspace{-3mm}
\end{figure}

As ODIs under each projection type are constrained by different transforming equations, the distortion caused by each type is inconsistent, indicating that applying matrix operations under one projection type can introduce unexpected changes when being converted to other types.
Specifically, applying uniformed bicubic downsampling on ERP images will affect the distribution of pixel density on Fisheye images, which are the original-formatted image type of imaging process in real-world scenarios.
To analyze the specific effect of ERP downsampling on the Fisheye image, we revisit the cause of distortions in ERP and Fisheye.
%

As we assume that viewing directions are uniformly distributed, the data points in an ideal ODI should be uniformly distributed on a spherical surface.
In practice, there is a trade-off between the uniformity of the spherical surface and the structural degree.
ERP is the most convenient projection type for storage or transmission, but it is also the projection type that suffers the heaviest distortion.
To better explain the causation of distortions, we follow the definition of stretching ratio ($\mathbf{K}$) in \cite{WSPSNR}, which represents distortion degree at different locations from the target projection type to the ideal spherical surface.
$\mathbf{K}$ is determined by area variation from one projection type to another.
When the target type is uniforming spherical surface, $\mathbf{K}$ is defined as:
\begin{equation}
	\mathbf{K}(x, y)=\frac{\delta S(\theta, \varphi)}{\delta P(x, y)}=\frac{\cos (\varphi)|d \theta d \varphi|}{|d x d y|}=\frac{\cos (\varphi)}{|J(\theta, \varphi)|},
\end{equation}
where $\delta S(\cdot, \cdot)$ and $\delta P(\cdot, \cdot)$ represent the area on the spherical surface and the projection plane, respectively.
$|d i d j|$ represents plane microunit.
$|J(\theta, \varphi)|$ is the Jacobian determinant from spherical coordinate to projection coordinate.
%

\textbf{ERP distortion.}
The coordinate in ERP is defined as $x=\theta$ and $y=\varphi$.
ERP stretching ratio can be derived as:
\begin{equation}\label{eq:k_erp}
	\mathbf{K}_{\operatorname{ERP}}(x, y)=\cos (\varphi)=\cos (y),
\end{equation}
where $x \in(-\pi, \pi)$, $y \in(-\frac{\pi}{2}, \frac{\pi}{2})$.
From \cref{eq:k_erp}, we conclude that ERP distortion is only determined by its latitude degree.
$\mathbf{K}_{\operatorname{ERP}}$ is reduced to zero when the absolute value of latitude degree increases to $\pi / 2$, which represents that pixel density on the polar areas of ERP images is closer to zero.
As shown in \cref{fig:erp_distortion} (c), with the increasing of the absolutely value of latitude degree ($|\varphi_s|$), the corresponding area on the spherical surface of an ERP microunit is gradually decreased to zero.
In conclusion, ERP distortion is caused by variable stretching ratios $\mathbf{K}_{\operatorname{ERP}}$, and is the heaviest in the polar areas.

\textbf{Fisheye distortion.}
The coordinate in Fisheye can be derived from $\theta=\arctan{(\frac{y}{x})}$ and $\varphi=(1-\sqrt{x^2+y^2})\times\frac{\pi}{2}$.
The stretching ratios of Fisheye can be derived as\footnote{Detailed derivative processes can be found in the supplementary file.}:
\begin{equation}\label{eq:k_fisheye}
	\mathbf{K}_{\operatorname{Fisheye}}(x, y)=\frac{\frac{2}{\pi} \sin{(\frac{\pi}{2} \sqrt{x^2+y^2})}}{\sqrt{x^2+y^2}}, 
\end{equation}
where $\sqrt{x^2+y^2} \in (0, 1)$.
$\mathbf{K}_{\operatorname{Fisheye}}$ is determined by distance from the fisheye center.
As $(\mathbf{K}_{\operatorname{Fisheye}})^{-1}$ is bounded, fisheye projection is closer to uniform distribution than ERP.
Moreover, it introduces much slighter distortion at the polar.

\textbf{Relationship between ERP and Fisheye distortions.}
To simplify, here we only discuss a typical Fisheye with an aperture degree of $\pi$ and a horizontal slicing plane\footnote{The influence of Fisheye formats with arbitrary splicing plane is discussed in the supplementary file.}.
In this case, the ERP coordinates and Fisheye's polar coordinates correspond linearly.
We can quantize the relationship by:
\begin{equation}\label{eq:k_r}
	\mathbf{K}_{\operatorname{ERP|Fisheye}}(\theta, \varphi)=\frac{\mathbf{K}_{\operatorname{ERP}}(x_E, y_E)}{\mathbf{K}_{\operatorname{Fisheye}}(x_F, y_F)}=\frac{\pi}{2}-|\varphi|,
\end{equation}
where $\theta, \varphi$ are spherical coordinates on the sphere, $x_E, y_E$ ($x_F, y_F$) denotes the plain coordinate under ERP (Fisheye).
From \cref{eq:k_erp,eq:k_fisheye,eq:k_r}, we conclude that when uniformed downsampling is performed on ERP, the kernel size of equivalent Fisheye downsampling is non-uniformed.
Especially when fisheye projection is spliced horizontally, the kernel size is proportional with $\pi/2 - |\varphi|$.

\subsection{Learning with More Realistic Degradation}\label{subs:fisheye_downsampling}

\begin{figure}[t]
	\centering
	\includegraphics[width=1\linewidth]{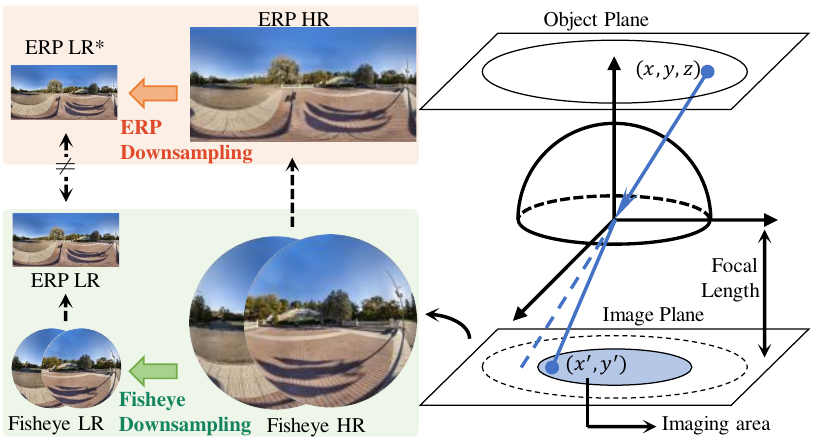}
	\vspace{-6mm}
	\caption{Downsampling process of ODIs (left) and imaging process in real world (right). * denotes that LR images synthesized from different downsampling processes are inconsistent.}
	\label{fig:imaging}
	\vspace{-5mm}
\end{figure}

\begin{figure*}[ht]
	\centering
	\vspace{-4mm}
	\resizebox{\textwidth}{!}{
	\includegraphics[]{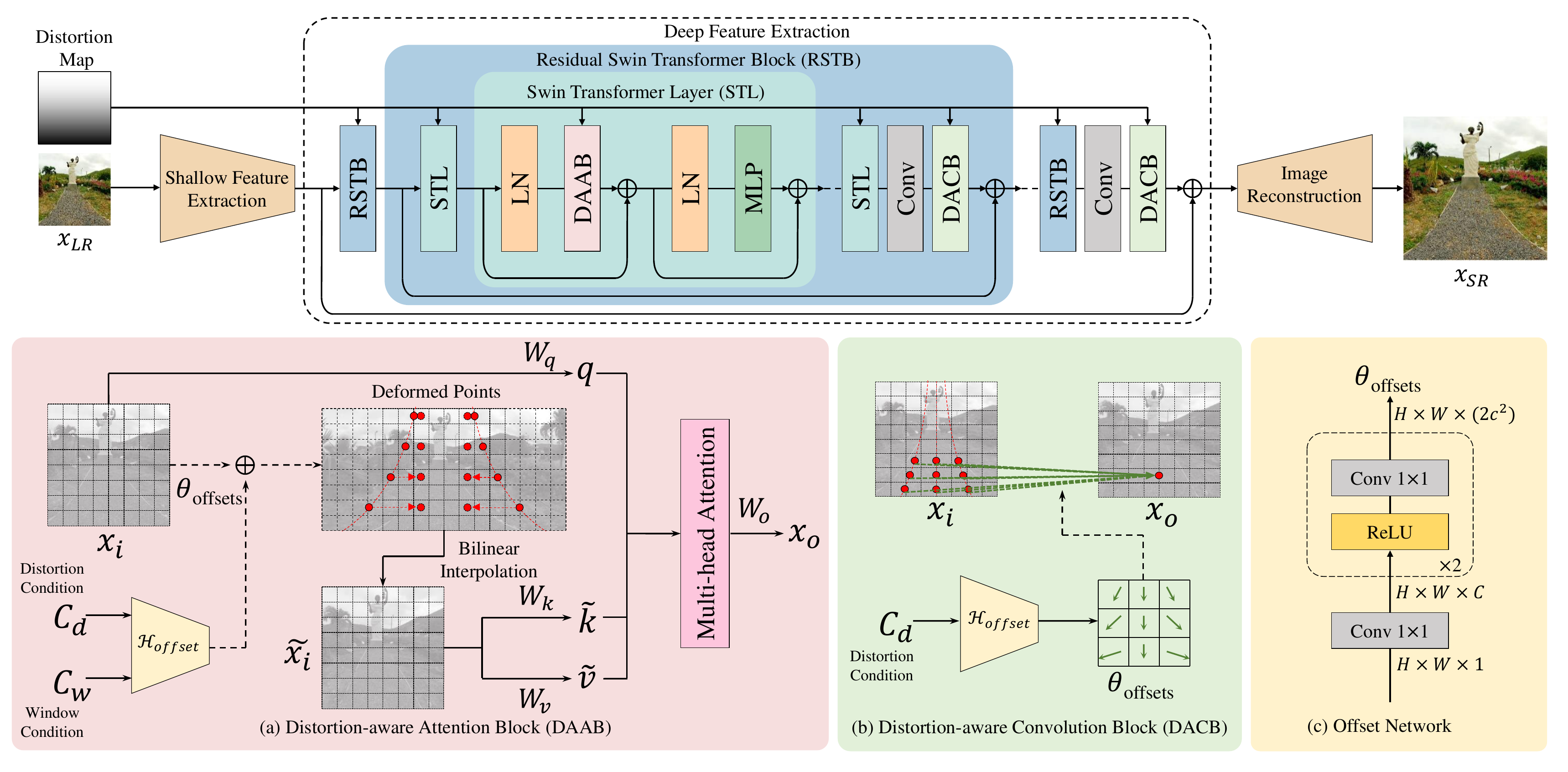}
}
	\vspace{-8mm}
	\caption{Overall illustration of OSRT. From SwinIR \cite{SwinIR}, we replace the standard multi-head self-attention block with DAAB and insert DACB behind the end of the RSTB. Channel dimensions of $\theta_{\operatorname{offsets}}$ in DAAB and DACB are 2 and 18, respectively.}
	\label{fig:backbone}
	\vspace{-6mm}
\end{figure*}

As depicted in \cref{fig:imaging}, the original-formatted projection type in ODI acquiring process is fisheye projection.
Given that real-world low-resolution issues are caused by insufficient precision and density of sensors, we consider that the degeneration process should be directly applied to original-formatted images before the type conversion.
Ideally, as camera sensors are arranged in uniform arrays, pixel density on original-formatted images is consistent everywhere.
Thus, for a realistic ODI, the pixel density on Fisheye should be a constant.
As discussed in \cref{subs:distortion}, applying uniformed downsampling on ERP means applying downsampling of variable kernel size on Fisheye.
The variable kernel size leads to variable Fisheye pixel density, which results in unrealistic LR images.
In conclusion, the ERP downsampling in previous methods influences the intrinsic distribution of pixel density in original-formatted images, which leads to unrealistic ODIs.
When the downsampling process happens on Fisheye, the Fisheye pixel density is unchanged, which fits the real-world imaging process and synthesizes more realistic LR pairs.

\textbf{Process of Fisheye downsampling.}
To generate more realistic LR ODIs, we mimic the real-world imaging process and apply bicubic downsampling on Fisheye images.
One single Fisheye image can only store information about a hemisphere. 
Hence, ERP images are converted to dual Fisheye images.
Before downsampling, Fisheye images are padded by a FOV larger than 180$^{\circ}$ to avoid edge disconnections.
This padding operation will not influence the geometric transforming relation between ERP and Fisheye.
As Fisheye data is unstructured and Fisheye distortion is more complicated than ERP distortion, we still learn the upscaling process under ERP.
Thus we reconvert LR images to the ERP format.
The overall process of Fisheye downsampling are described in \cref{fig:imaging}.

\subsection{OSRT: Modulate Distortion in ODIs}\label{subs:OSRT}

\textbf{Overall.}
As discussed in \cref{subs:distortion}, ERP images suffer a distortion caused by a non-consistency area stretching ratio from an ideal spherical surface.
Referred from \cref{eq:k_erp}, for an LR input $X_i \in \mathbb{R}^{C\times M\times N}$, the distortion map $C_d \in \mathbb{R}^{1\times M\times N}$ is derived by: 
\begin{equation}\label{eq:condition}
	C_d = \cos \left(\frac{m+0.5-M / 2}{M} \pi\right),
\end{equation}
where $m$ is the current height of LR input.

Previous methods tend to treat $C_d$ as an additional input of $X_i$ \cite{DistortedPers}, or re-weighting parameters by $C_d$ \cite{GAConv}.
Although these solutions can benefit from building awareness of distortion, continuous and amorphous distortions cannot be adequately fitted by scattering and structured convolution operations.
While previous methods cannot fully explore the advantage of $C_d$, we intend to design a novel block for learning distorted patterns continuously.
In VSR tasks, the deformable mechanism is proposed to align features between adjacent frames \cite{TDAN,EDVR}.
Unlike standard DCN \cite{DCNv1}, which calculates offsets from the input feature map, offsets are calculated from bi-directional optical flow in VSR pipelines.
Inspired by feature-level flow warping in VSR, we find that the deformable mechanism is a feasible solution for continuous mappings.
Consequently, we modulate ERP distortion by feature-level warping operations.
As shown in \cref{fig:backbone}, $C_d$ is only utilized to calculate the deformable offsets $\Delta p$.
To keep compatibility with arbitrary ConvNets and Transformers, we propose two blocks to modulate ERP distortion, which can directly replace the multi-head self-attention blocks in Transformers and the standard convolution layers in ConvNets, respectively.

\textbf{Distortion-aware attention block (DAAB).}
As depicted in \cref{fig:backbone} (a), a distortion condition guided deformable self-attention is proposed to learn correlations between the distorted input $F_{i-1}$ and its corresponding modulated feature map $\tilde{F}_{i-1}$.
DAAB is formulated as:
\begin{equation}
	\Delta p_i=H_{\operatorname{offset}_i}(C_d, C_w), \tilde{F}_{i-1}=\phi(F_{i-1} ; p_i+\Delta p_i),
\end{equation}
\begin{equation}
	F_i = H_{\operatorname{SA}}(F_{i-1} W_{q_i},\tilde{F}_{i-1} W_{k_i},\tilde{F}_{i-1} W_{v_i}),
\end{equation}
where $H_{\operatorname{offset}_i}(\cdot)$ denotes the $i$-th convolution block to calculate offset maps $\Delta p_i \in \mathbb{R}^{2\times H\times W}$, and $H_{\operatorname{SA}}$ denotes standard self-attention formula.
$H_{\operatorname{offset}}(\cdot)$ consists of $1\times1$ convolution block with two hidden layers.
The input of $H_{\operatorname{offset}}(\cdot)$ is concatenated by the latitude-related distortion condition $C_d \in \mathbb{R}^{1\times H\times W}$ and the window condition $C_w \in \mathbb{R}^{2\times H\times W}$.
$C_w$ is a linear position encoding within a self-attention kernel.
$\phi(\cdot, \cdot)$ denotes a bilinear interpolation, and $W_{q_i}, W_{k_i}, W_{v_i}$ denote $i$-th weight matrix of query, key, and value, respectively.
For multi-head self-attention blocks, $H_{\operatorname{offset}_i}(\cdot)$ is identical in calculations of parallel heads.

\textbf{Distortion-aware convolution block (DACB).}
As shown in \cref{fig:backbone} (b), we apply a standard deformable convolution layer with a substituted input for offset calculation.
Modulated output $F_i$ is extracted as:
\begin{equation}
	\Delta p_i=H_{\operatorname{offset}_i}(C_d),F_i=H_{\operatorname{DCN}_i}(F_{i-1}, \Delta p_i),
\end{equation}
where $H_{\operatorname{DCN}}(F, \Delta p)$ denotes standard deformable convolution layer in \cite{DCNv2}.
The architecture of $H_{\operatorname{offset}_i}(\cdot)$ is identical to that in DAAB.
As the kernel size of DCN is $3\times3$ in DACB, the output channel dimension of offsets maps is 18.

\textbf{OSRT.}
In practice, we propose an Omnidirectional image Super-Resolution Transformer, named OSRT.
SwinIR \cite{SwinIR} is selected as the basic architecture for its strong reconstruction ability in the SISR task.
To learn distortion rectified representations, we stack a DACB after the last convolution layer of each residual swin Transformer block and replace all self-attention blocks as DAAB.
The feature dimension of OSRT is reduced from 180 to 156 to maintain identical parameters with SwinIR.

\begin{figure}[b]
	\centering
	\vspace{-4mm}
	\includegraphics[width=1\linewidth]{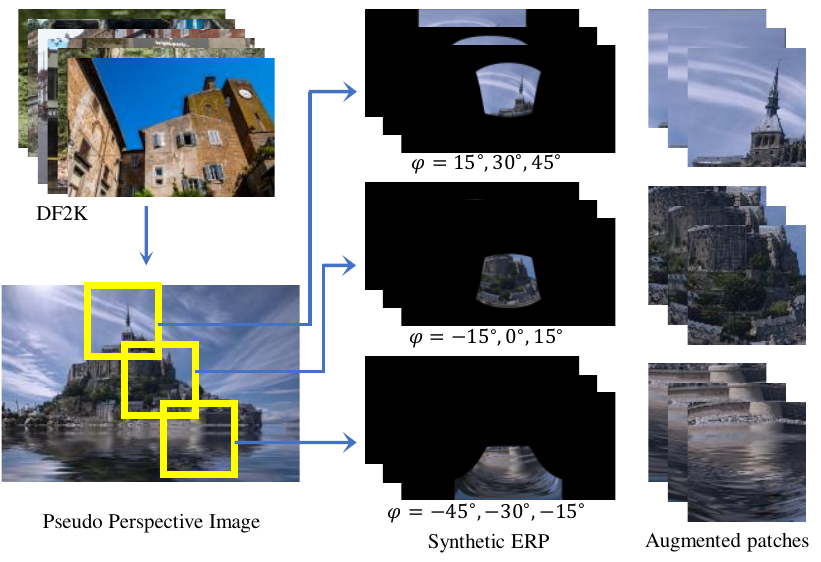}
	\vspace{-6mm}
	\caption{Synthetic process of DF2K-ERP.}
	\label{fig:distortion_aug}
\end{figure}

\subsection{Boosting ODISR Performance by Plain Images}\label{subs:dataaug}

As the capacity of OSRT is relatively large, it suffers overfitting for large upscaling factors (\cref{fig:overfit}).
Given that acquiring ODIs are expensive, we propose to generate pseudo ERP images from 2D plain images to tackle this issue.
After being sampled by sliding windows, the patch of plain images is treated as a plain perspective.
By converting from Perspective to ERP, plain images are distorted in the same way as ERP.
Considering that distortion of a Perspective is enlarged by its FOV degree, a relatively small FOV degree of 90$^{\circ}$ is applied. 
For a given pseudo Perspective, $\theta_p$ is fixed at $0$ and $\varphi_p$ is derived by:
\begin{equation}
	\varPhi_p = \varphi_h + z_0,
\end{equation}
where $\varphi_h$ is determined by patch locations and $z_0$ is orderly sampled from $\{-15^{\circ}, 0^{\circ}, 15^{\circ}\}$.
To maximize the approximate data distribution of ODIs, we horizontally split a plain image into three sub-images and define $\varphi_h$ as $-30^{\circ}, 0^{\circ}, 30^{\circ}$ respectively.
Pseudo ERP images are cropped to remove the black border.
As shown in \cref{fig:distortion_aug}, we get a new ERP dataset (called DF2K-ERP) by implementing the augmentation pipeline on widely-used plain image dataset DF2K \cite{DIV2K,EDSR}.
The DF2K-ERP dataset consists of 146K high-quality ERP image patches with a patch size larger than 256.

\begin{table*}[h]
	\centering
	\scriptsize
		\begin{tabular}{c|c|p{0.058\textwidth}<{\centering}p{0.058\textwidth}<{\centering}|cc|p{0.058\textwidth}<{\centering}p{0.058\textwidth}<{\centering}|cc}
			\toprule
			\multicolumn{1}{c|}{\multirow{2}{*}{Method}}             & \multirow{2}{*}{Scale} & \multicolumn{4}{c|}{ODI-SR}         & \multicolumn{4}{c}{SUN 360 Panorama} \\
			\multicolumn{1}{c|}{}&                                   & PSNR  & \multicolumn{1}{c}{SSIM}   & WS-PSNR & WS-SSIM & PSNR   & \multicolumn{1}{c}{SSIM}    & WS-PSNR & WS-SSIM \\ \midrule
			Bicubic                             & \multirow{7}{*}{$\times$2} & 28.21 & 0.8215 & 27.61   & 0.8156  & 28.14  & 0.8118  & 28.01   & 0.8321  \\ 
			RCAN \cite{RCAN}                    &                        & 30.08 & 0.8723 & 29.49   & 0.8714  & 30.56  & 0.8712  & 31.18   & 0.8969  \\ 
			SRResNet \cite{ESRGAN}              &               & 30.16 & 0.8717 & 29.59   & 0.8697  & 30.65  & 0.8714  & 31.20   & 0.8953  \\ 
			EDSR \cite{EDSR}                    &              & 30.32 & 0.8770 & 29.68   & 0.8727  & 30.89  & 0.8784  & 31.42   & 0.8995  \\ 
			SwinIR \cite{SwinIR}                &               & 30.52 & 0.8819 & 29.87   & 0.8772  & 31.21  & 0.8852  & 31.78   & 0.9051  \\ 
			SwinIR$^{\dagger}$ \cite{SwinIR}    &              & 30.64 & 0.8821 & 30.00   & 0.8777  & 31.33  & 0.8855  & 31.98   & 0.9059  \\ 
			OSRT$^{\dagger}$                    &              & \textbf{30.77} & \textbf{0.8846} & \textbf{30.11} & \textbf{0.8795} & \textbf{31.52} & \textbf{0.8888} & \textbf{32.14} & \textbf{0.9081}   \\ \midrule
			Bicubic                             & \multirow{7}{*}{$\times$4}              & 25.59 & 0.7118 & 24.95   & 0.6923  & 25.29  & 0.6993  & 24.90   & 0.7083  \\ 
			RCAN \cite{RCAN}                    &               & 26.85 & 0.7621 & 26.15   & 0.7485  & 27.10  & 0.7660  & 26.99   & 0.7856  \\ 
			SRResNet \cite{ESRGAN}              &              & 26.91 & 0.7597 & 26.24   & 0.7457  & 27.10  & 0.7618  & 26.99   & 0.7812  \\ 
			EDSR \cite{EDSR}                    &              & 26.97 & 0.7589 & 26.30   & 0.7458  & 27.19  & 0.7633  & 27.10   & 0.7827  \\ 
			SwinIR \cite{SwinIR}                &              & 27.12 & 0.7663 & 26.44   & 0.7523  & 27.39  & 0.7707  & 27.30   & 0.7901  \\
			SwinIR$^{\dagger}$ \cite{SwinIR}    &              & 27.31 & 0.7735 & 26.61   & 0.7589  & 27.71  & 0.7804  & 27.64   & 0.7996  \\ 
			OSRT$^{\dagger}$                    &              & \textbf{27.41} & \textbf{0.7762} & \textbf{26.70} & \textbf{0.7609} & \textbf{27.84} & \textbf{0.7835} & \textbf{27.77} & \textbf{0.8020}  \\ \bottomrule
		\end{tabular}
	\vspace{-2mm}
	\caption{SR results under Fisheye downsampling. $\dagger$ denotes applying DF2K-ERP as augmented dataset. Best results are shown in \textbf{Bold}.}
	\label{tab:main}
	\vspace{-2mm}
\end{table*}

\begin{figure*}[h]
	\scriptsize
	\centering
	\begin{tabular}{l}
		\hspace{-0.42cm}
		\begin{adjustbox}{valign=t}
			\begin{tabular}{c}
				\includegraphics[width=0.260\textwidth]{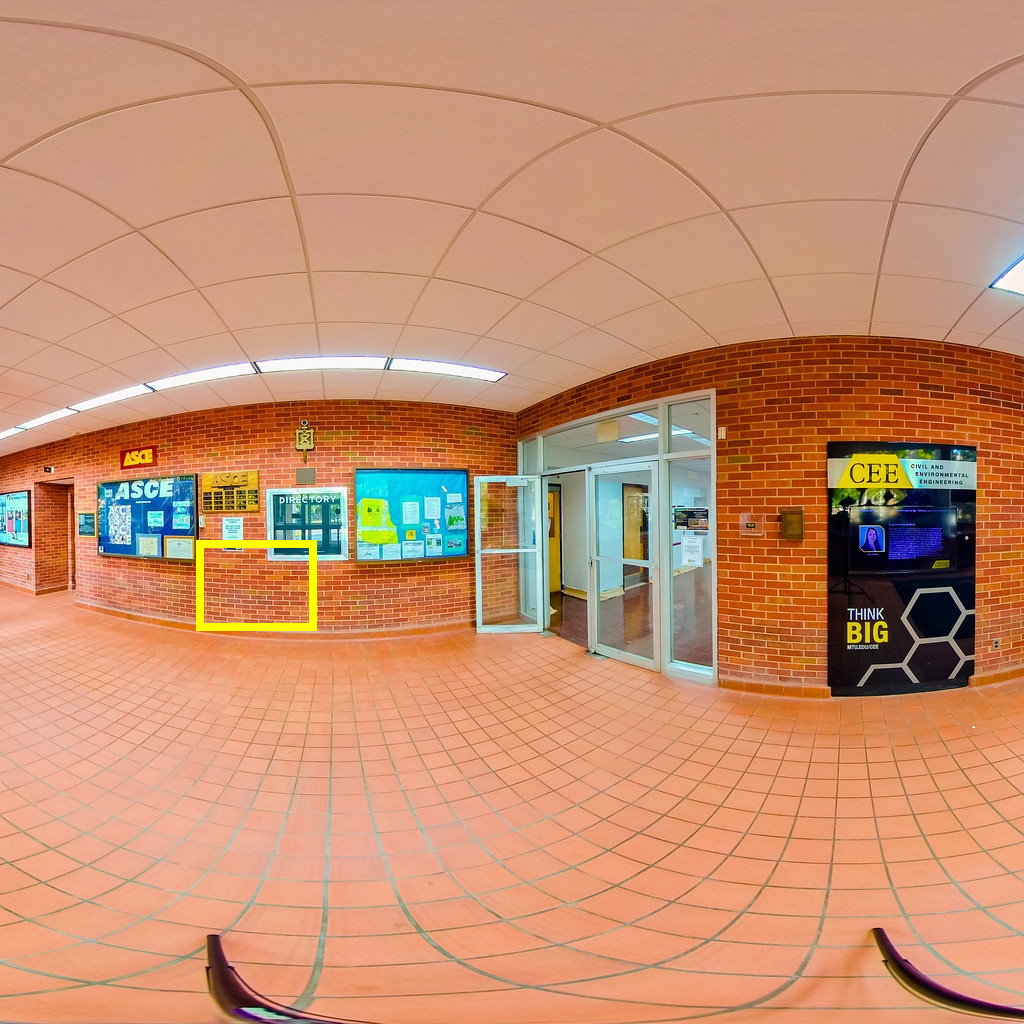}
				\vspace{1.5mm}
				\\
				SUN360 ($\times$4): 034
			\end{tabular}
		\end{adjustbox}
		\hspace{-2mm}
		\begin{adjustbox}{valign=t}
			\begin{tabular}{cccc}
				\includegraphics[width=0.149\textwidth]{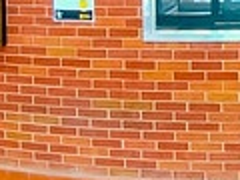} \hspace{-1mm} &
				\includegraphics[width=0.149\textwidth]{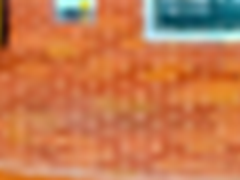} \hspace{-1mm} &
				\includegraphics[width=0.149\textwidth]{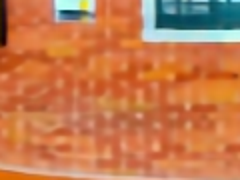} \hspace{-1mm} &
				\includegraphics[width=0.149\textwidth]{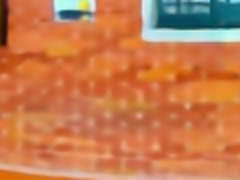} \hspace{-1mm} 
				\\
				HR \hspace{-1mm} &
				Bicubic \hspace{-1mm} &
				RCAN \cite{RCAN} \hspace{-1mm} &
				SRResNet \cite{ESRGAN} \hspace{-1mm} 
				\\
				PSNR/SSIM \hspace{-1mm} &
				24.38dB/0.6872 \hspace{-1mm} &
				26.40dB/0.8137 \hspace{-1mm} &
				26.21dB/0.7999 \hspace{-1mm} 
				\\
				\includegraphics[width=0.149\textwidth]{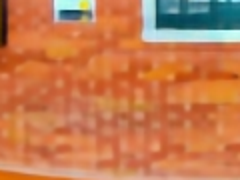} \hspace{-1mm} &
				\includegraphics[width=0.149\textwidth]{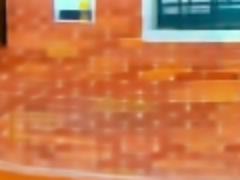} \hspace{-1mm} &
				\includegraphics[width=0.149\textwidth]{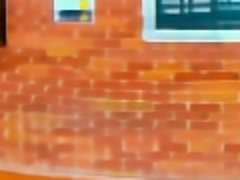} \hspace{-1mm} &
				\includegraphics[width=0.149\textwidth]{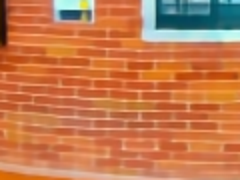} \hspace{-1mm}  
				\\ 
				EDSR \cite{EDSR} \hspace{-1mm} &
				SwinIR \cite{SwinIR} \hspace{-1mm} &
				SwinIR$^{\dagger}$ \cite{SwinIR} \hspace{-1mm} &
				OSRT$^{\dagger}$ \hspace{-1mm} 
				\\
				26.38dB/0.8072 \hspace{-1mm} &
				26.77dB/0. 8234 \hspace{-1mm} &
				27.34dB/0.8462 \hspace{-1mm} &
				27.68dB/0.8561 \hspace{-1mm} 
			\end{tabular}
		\end{adjustbox}
		\vspace{2mm}
		
		\\ 
		\hspace{-0.42cm}
		\begin{adjustbox}{valign=t}
			\begin{tabular}{c}
				\includegraphics[width=0.260\textwidth]{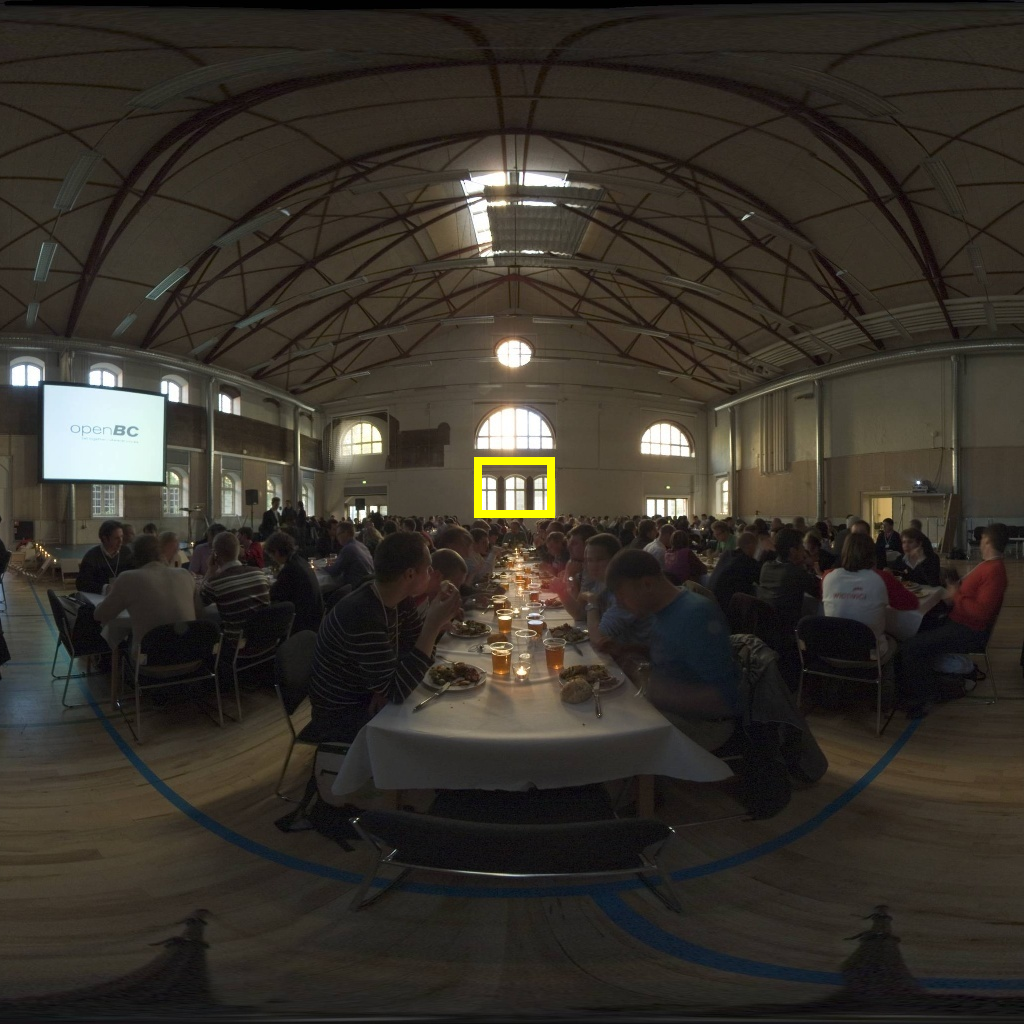}
				\vspace{1.5mm}
				\\
				SUN360 ($\times$4): 095
			\end{tabular}
		\end{adjustbox}
		\hspace{-2mm}
		\begin{adjustbox}{valign=t}
			\begin{tabular}{cccc}
				\includegraphics[width=0.149\textwidth]{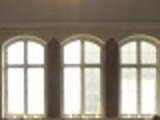} \hspace{-1mm} &
				\includegraphics[width=0.149\textwidth]{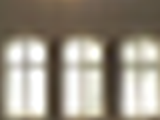} \hspace{-1mm} &
				\includegraphics[width=0.149\textwidth]{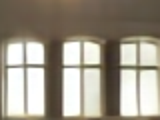} \hspace{-1mm} &
				\includegraphics[width=0.149\textwidth]{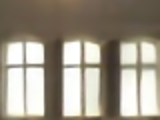} \hspace{-1mm} 
				\\
				HR \hspace{-1mm} &
				Bicubic \hspace{-1mm} &
				RCAN \cite{RCAN} \hspace{-1mm} &
				SRResNet \cite{ESRGAN} \hspace{-1mm} 
				\\
				PSNR/SSIM \hspace{-1mm} &
				30.20dB/0.8506 \hspace{-1mm} &
				33.59dB/0.9088 \hspace{-1mm} &
				33.43dB/0.9043 \hspace{-1mm} 
				\\
				\includegraphics[width=0.149\textwidth]{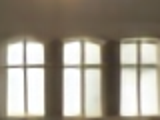} \hspace{-1mm} &
				\includegraphics[width=0.149\textwidth]{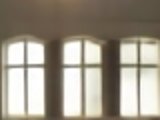} \hspace{-1mm} &
				\includegraphics[width=0.149\textwidth]{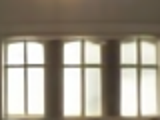} \hspace{-1mm} &
				\includegraphics[width=0.149\textwidth]{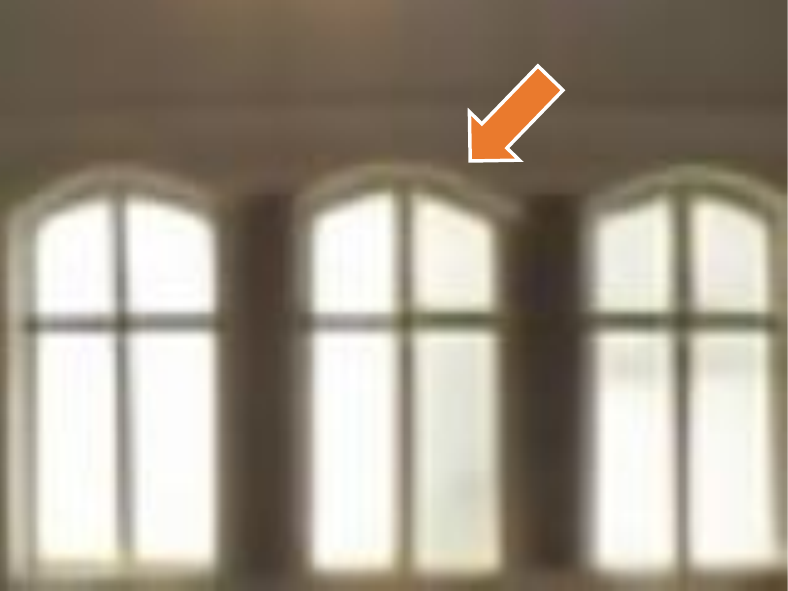} \hspace{-1mm}  
				\\ 
				EDSR \cite{EDSR} \hspace{-1mm} &
				SwinIR \cite{SwinIR} \hspace{-1mm} &
				SwinIR$^{\dagger}$ \cite{SwinIR} \hspace{-1mm} &
				OSRT$^{\dagger}$ \hspace{-1mm} 
				\\
				33.64dB/0.9074 \hspace{-1mm} &
				34.05dB/0.9119 \hspace{-1mm} &
				34.41dB/0.9158 \hspace{-1mm} &
				34.77dB/0.9187 \hspace{-1mm} 
			\end{tabular}
		\end{adjustbox}
		\\ 
		
	\end{tabular}
	\vspace{-3mm}
	\caption{Visual comparisons of $\times$4 SR results under Fisheye downsampling.}
	\label{fig:main_x4_visual}
	\vspace{-4mm}
\end{figure*}

\section{Experiments}
\label{sec:Exp}

\subsection{Experimental Setup}
ODI-SR dataset \cite{LauNet} and SUN360 Panorama dataset \cite{Sun360} are used in our experiment.
In the training phase, we follow the data split setting in \cite{LauNet} and train on the ODI-SR training set.
The resolution of the ERP HR is $1024 \times 2048$, and the upscaling factors are $\times2$ and $\times4$.
Fisheye downsampling is applied as our pre-defined downsampling kernel.
Loss is calculated by L1 distance and optimized by Adam \cite{ADAM}, with an initial learning rate of $2\times10^{-4}$, a total batch size of 32, and an input patch size of $64$.
We train OSRT for $500k$ iterations and halve the learning rate at $250k$, $400k$, $450k$ and $475k$.  
In evaluation, we test on the ODI-SR testing set and SUN360 dataset.
PSNR \cite{PSNR}, SSIM \cite{SSIM}, and their distortion re-weighted versions (WS-PSNR \cite{WSPSNR}, WS-SSIM \cite{WSSSIM}) are used as evaluation metrics. 

\begin{table*}[h]
	\centering
	\scriptsize
	\begin{tabular}{c|cc|cc|cc|cc}
		\toprule
		Scale & \multicolumn{4}{c|}{$\times$8} & \multicolumn{4}{c}{$\times$16} \\ \midrule
		\multirow{2}{*}{Method} & \multicolumn{2}{c|}{ODI-SR} & \multicolumn{2}{c|}{SUN 360 Panorama} & \multicolumn{2}{c|}{ODI-SR} & \multicolumn{2}{c}{SUN 360 Panorama} \\
		& WS-PSNR & WS-SSIM & WS-PSNR & WS-SSIM & WS-PSNR & WS-SSIM & WS-PSNR & WS-SSIM \\ \midrule
		Bicubic & 19.64 & 0.5908 & 19.72 & 0.5403 & 17.12 & 0.4332 & 17.56 & 0.4638 \\ 
		SRCNN \cite{SRCNN} & 20.08 & 0.6112 & 19.46 & 0.5701 & 18.08 & 0.4501 & 17.95 & 0.4684 \\ 
		EDSR \cite{EDSR} & 23.97 & 0.6417 & 22.46 & 0.6341 & 21.12 & 0.5698 & 21.06 & 0.5645 \\ 
		RCAN \cite{RCAN} & 24.26 & 0.6628 & 23.88 & 0.6542 & 21.94 & 0.5824 & 21.74 & 0.5742 \\ 
		360-SS \cite{360SS} & 21.65 & 0.6417 & 21.48 & 0.6352 & 19.65 & 0.5431 & 19.62 & 0.5308 \\ 
		LAU-Net \cite{LauNet} & 24.36 & 0.6801 & 24.02 & 0.6708 & 22.07 & 0.5901 & 21.82 & 0.5824 \\ 
		SphereSR \cite{SphereSR} & 24.37 & 0.6777 & 24.17 & 0.6820 & 22.51 & \textbf{0.6370} & 21.95 & 0.6342 \\ 
		OSRT & \textbf{24.53} & \textbf{0.6780} & \textbf{24.38} & \textbf{0.7072} & \textbf{22.69} & 0.6261 & \textbf{22.13} & \textbf{0.6388} \\ \bottomrule
	\end{tabular}
	\vspace{-2mm}
	\caption{SR results under ERP downsampling.}
	\label{tab:erp_setting}
	\vspace{-5mm}
\end{table*}

\subsection{Evaluation under Fisheye Downsampling}

When the downsampling process is performed on Fisheye images, we train SRResNet \cite{ESRGAN}, EDSR \cite{EDSR}, RCAN \cite{RCAN}, and SwinIR \cite{SwinIR} for comparison.

\textbf{Quantitative results.}
As shown in \cref{tab:main}, with the help of additional DF2K-ERP training patches, OSRT outperforms previous methods by $~$0.3dB on PSNR.
Although directly applying SwinIR on the ODISR task has already reached SOTA performance, OSRT surpasses SwinIR over 0.1dB on two datasets for both $\times$2 and $\times$4 SR tasks, which demonstrates the effectiveness of its distortion modulation ability.
The performance of RCAN degrades under Fisheye downsampling, which is caused by the incompatibility between channel attention and Fisheye downsampling\footnote{The cause is discussed in the supplementary file.}. 

\begin{figure}[b]
	\vspace{-6mm}
	\scriptsize
	\centering
	\begin{tabular}{l}
		\hspace{-0.42cm}
		\begin{adjustbox}{valign=t}
			\begin{tabular}{c}
				\includegraphics[width=0.46\columnwidth]{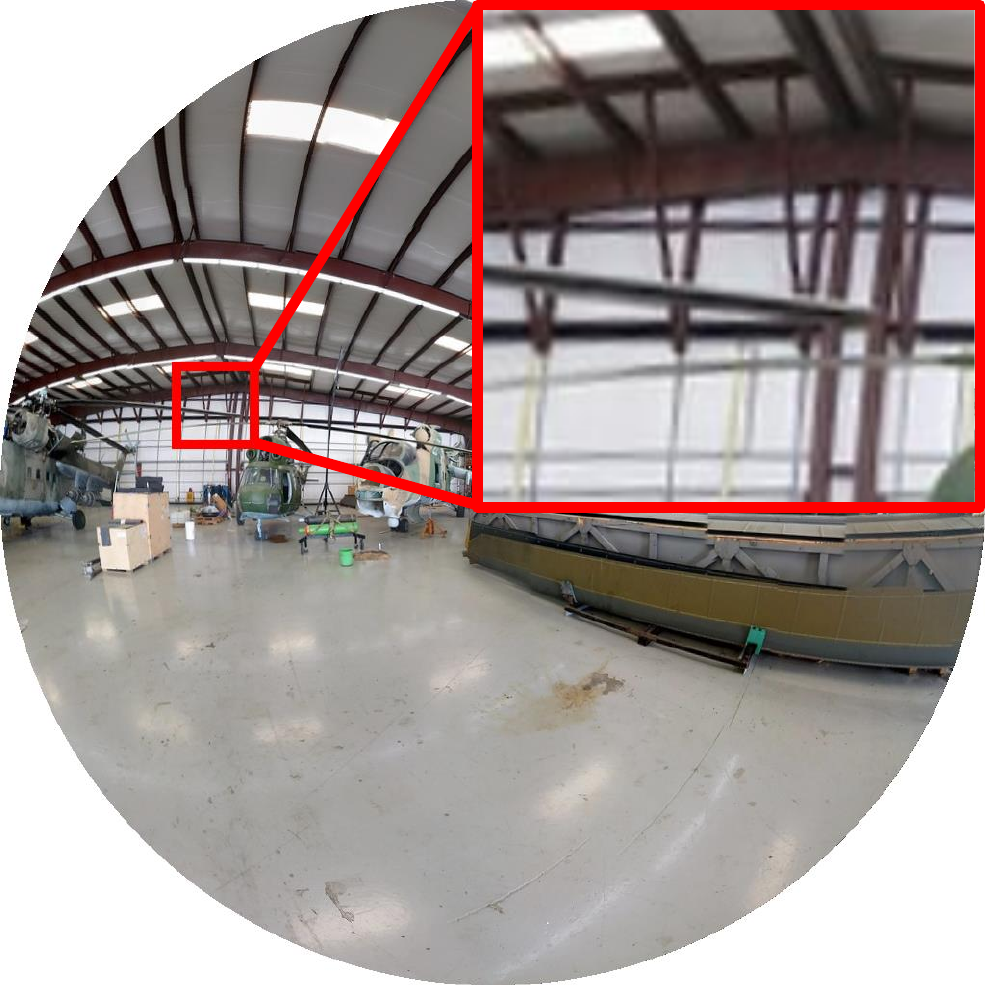}
				\\
				ODI-SR ($\times$4): 008
				\\
				Fisheye (Vertical, Left)
			\end{tabular}
		\end{adjustbox}
		\hspace{-3mm}
		\begin{adjustbox}{valign=t}
			\begin{tabular}{cc}
				\includegraphics[width=0.23\columnwidth]{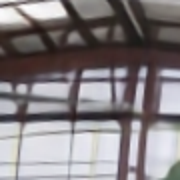} \hspace{-3mm} &
				\includegraphics[width=0.23\columnwidth]{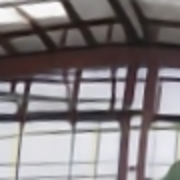} \hspace{-1mm} 
				\\
				EDSR \cite{EDSR} \hspace{-3mm} &
				SwinIR \cite{SwinIR} \hspace{-1mm} 
				\\
				\includegraphics[width=0.23\columnwidth]{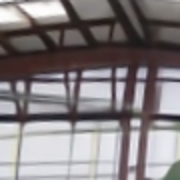} \hspace{-3mm} &
				\includegraphics[width=0.23\columnwidth]{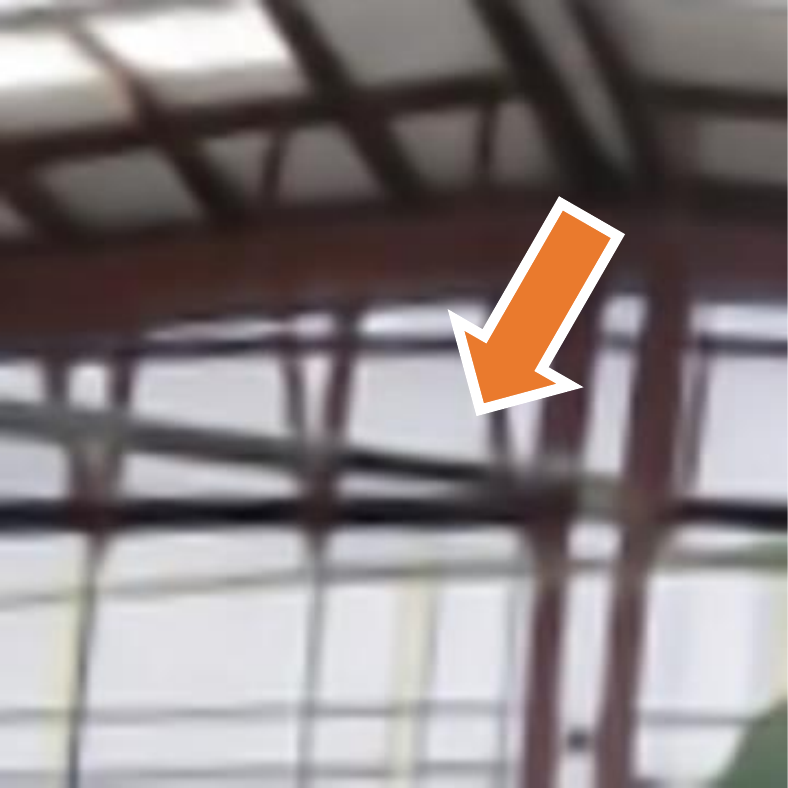} \hspace{-1mm} 
				
				\\ 
				SwinIR$^{\dagger}$ \cite{SwinIR} \hspace{-3mm} &
				OSRT$^{\dagger}$ \hspace{-1mm} 
				
			\end{tabular}
		\end{adjustbox}
		\vspace{1mm}
		
		\\ 
		\hspace{-0.42cm}
		\begin{adjustbox}{valign=t}
			\begin{tabular}{c}
				\includegraphics[width=0.46\columnwidth]{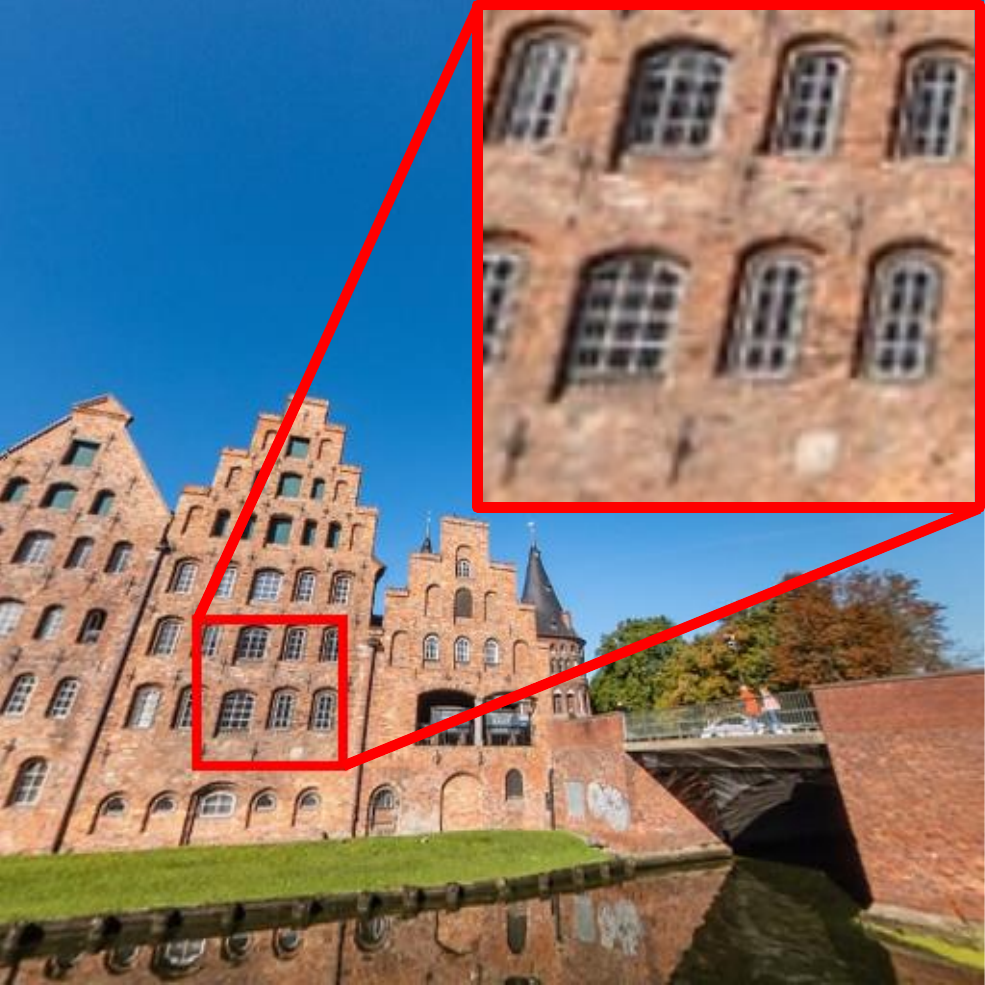}
				\\
				SUN360 ($\times$2): 062
				\\
				Perspective ($\varphi$: $30^{\circ}$; FOV: $90^{\circ}$)
			\end{tabular}
		\end{adjustbox}
		\hspace{-3mm}
		\begin{adjustbox}{valign=t}
			\begin{tabular}{cc}
				\includegraphics[width=0.23\columnwidth]{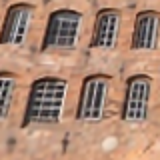} \hspace{-3mm} &
				\includegraphics[width=0.23\columnwidth]{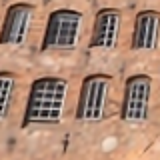} \hspace{-1mm} 
				\\
				EDSR \cite{EDSR} \hspace{-3mm} &
				SwinIR \cite{SwinIR} \hspace{-1mm} 
				\\
				\includegraphics[width=0.23\columnwidth]{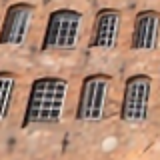} \hspace{-3mm} &
				\includegraphics[width=0.23\columnwidth]{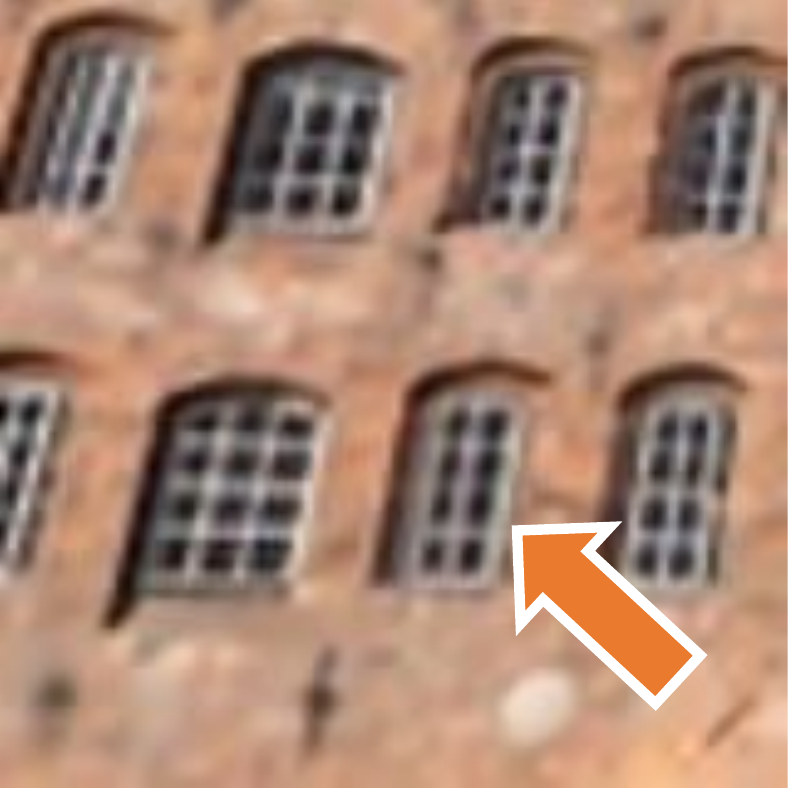} \hspace{-1mm} 
				
				\\ 
				SwinIR$^{\dagger}$ \cite{SwinIR} \hspace{-3mm} &
				OSRT$^{\dagger}$ \hspace{-1mm} 
				
			\end{tabular}
		\end{adjustbox}
		
	\end{tabular}
	\vspace{-3mm}
	\caption{Visual comparisons for SR of Fisheye and Perspective images. $\dagger$ denotes applying DF2K-ERP as augmented dataset.}
	\label{fig:main_non_erp}

\end{figure}

\textbf{Qualitative comparison.}
\cref{fig:main_x4_visual} shows the visualization results of $\times4$ ODISR task.
While other methods struggle to understand the geometric transformation process in distorted images, OSRT can reconstruct sharp and accurate boundaries with the advantages of distortion modulation.
It is observed that OSRT is skilled at reconstructing rigid texture.
Moreover, benefiting from the distortion modulation ability, OSRT can preserve the original structure as most when being projected to other projection types (\cref{fig:main_non_erp}).

\subsection{Evaluation under ERP Downsampling}

To compare with previous ODISR methods \cite{360SS,LauNet,SphereSR}, we train OSRT under the previous ERP setting.
Regardless of over-fitting issues, we only train on the dataset provided by \cite{LauNet} for fairness.
As shown in \cref{tab:erp_setting}, OSRT still outperforms LAU-Net \cite{LauNet} and SphereSR \cite{SphereSR} under large upscaling factor and ERP downsampling.
Without a complicated training pipeline and discrete inference process, OSRT yields the best PSNR values and surpasses all previous methods on most SSIM-related metrics (three of four). 

\subsection{Ablation Study and Discussion}

In this section, we prove the effectiveness of Fisheye downsampling, OSRT components, and augmented DF2K-ERP.
We then explain the distortion modulation ability of OSRT by visualizing offsets in deformable blocks.

\textbf{Fisheye downsampling.}
As shown in \cref{fig:intro_x8_visual}, the SR model trained under ERP downsampling is more likely to generate blur details and missing structures in real-world scenarios.
These artifacts cannot be removed by a superior backbone network, but can be eliminated by a more realistic imaging process.
More importantly, ERP downsampling directly covers the geometric property of ERP images and makes the ODISR task identical to the standard plain image super-resolution task.
The evidence is that a standard SISR model (SwinIR) trained on a plain image dataset (DF2K) can outperform previous SOTA in the ODISR task, which yields WS-PSNR results of 24.63dB/24.49dB (22.68dB/22.13dB) on $\times$8 ($\times$16) ODI-SR/SUN360 testing set, respectively. 
In conclusion, when the intrinsic property of ODIs is broken by ERP downsampling, the ODISR task degenerates into a plain image super-resolution task with a particular data distribution.

\begin{table}[t]
	\centering
	\scriptsize
	\setlength{\tabcolsep}{1.5mm}{
		\begin{tabular}{c|c|c|cc|cc|c}
			\toprule
			feature & \multirow{2}{*}{DACB} & \multirow{2}{*}{DAAB} & \multicolumn{2}{c|}{ODI-SR} & \multicolumn{2}{c|}{SUN360} & Params. \\ 
			dim &  &  & PSNR & SSIM & PSNR & SSIM & (M) \\ \midrule
			60 & $\times$ &  $\times$ & 30.27 & 0.8739 & 30.78 & 0.8742 & 0.91 \\ 
			60 & \checkmark & $\times$ & 30.41 & 0.8775 & 31.00 & 0.8793 & 1.16 \\ 
			60 & $\times$ & w/o $C_w$ & 30.31 & 0.8746 & 30.83 & 0.8755 & 1.00 \\ 
			60 & $\times$ & w/ $C_w$ & 30.32 & 0.8746 & 30.84 & 0.8753 & 1.01 \\
			60 & \checkmark & w/ $C_w$ &  \textbf{30.44} & \textbf{0.8780} & \textbf{31.04} & \textbf{0.8800} & 1.26 \\ 
			72 & $\times$ & $\times$ & 30.32 & 0.8748 & 30.85 & 0.8755 & 1.29 \\ \bottomrule
	\end{tabular}
}
	\vspace{-2mm}
	\caption{Ablation study on OSRT components. All models are trained on $\times2$ SR task under Fisheye downsampling.}
	\label{tab:archi}
	\vspace{-7mm}
\end{table}

\textbf{OSRT components.}
To study the effectiveness of each component in OSRT, we propose a light version of OSRT (OSRT-light) for ablation study, which corresponds with the official SwinIR-light \cite{SwinIR}.
As proofed in \cref{tab:archi}, all components in OSRT are beneficial for modulating ERP distortion.
The advantages of DACB and DAAB can be stacked when being applied in the same network.
Compared with simply expanding the feature dimension of SwinIR to match the network complexity, the overall improvements of OSRT is more significant (+0.05dB vs. +0.2dB).

\begin{figure}[t]
	\centering
	\includegraphics[width=1\linewidth]{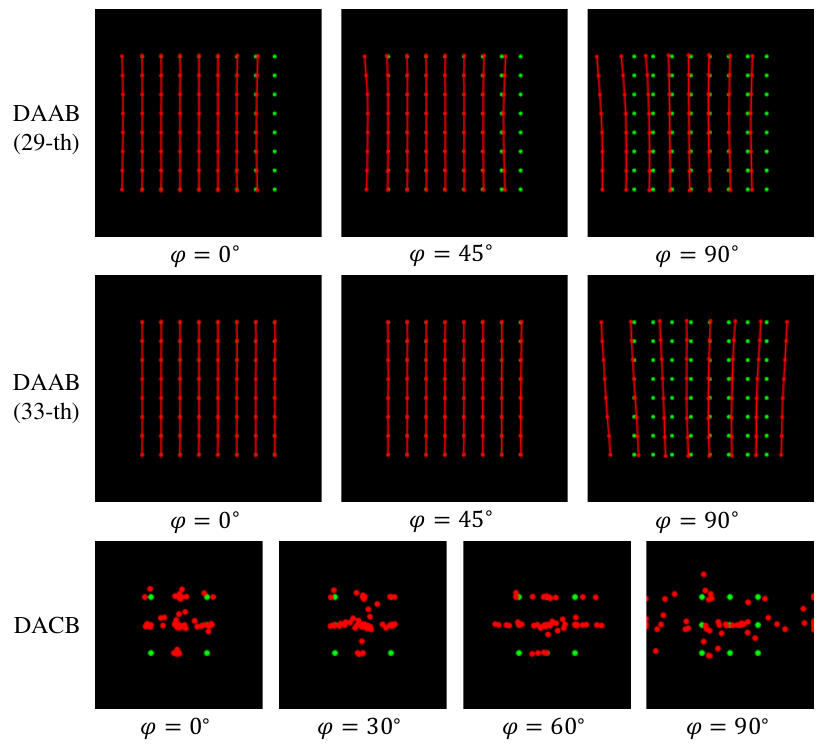}
	\vspace{-6mm}
	\caption{Visualizations of offset maps in OSRT. Reference and deformed points are depicted in \textcolor{green}{green} and \textcolor{red}{red}, respectively. The deformable kernel is sparse in the polar area.}
	\label{fig:offsets_visual}
	\vspace{-6mm}
\end{figure}

\begin{figure}[b]
	\vspace{-5mm}
	\centering
	\includegraphics[width=1\linewidth]{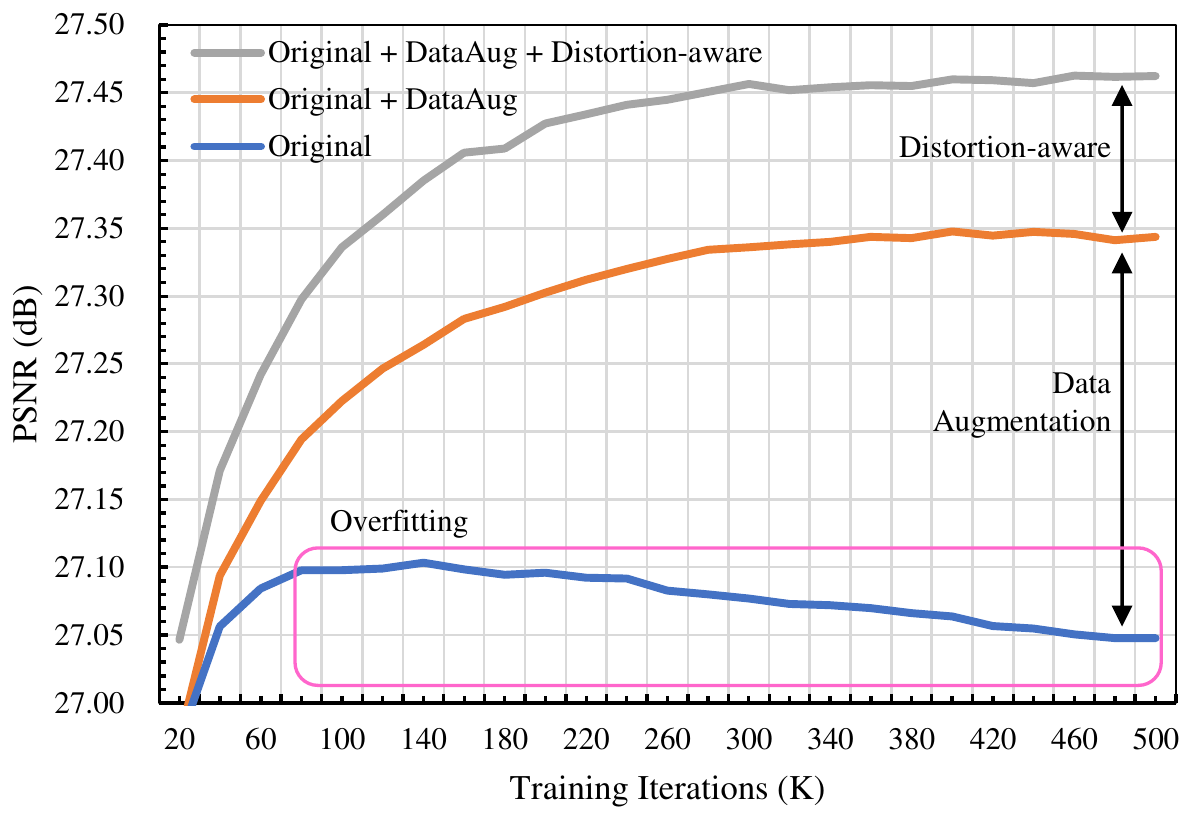}
	\vspace{-7mm}
	\caption{Training process of Transformers on $\times$4 ODISR task. The overfitting issue is tackled by our augmentation scheme.}
	\label{fig:overfit}
\end{figure}

\textbf{Offsets in OSRT.}
To investigate whether the deformable mechanism in OSRT can modulate distortion as expected, we visualize offsets map in a well-trained OSRT.
As depicted in \cref{fig:offsets_visual}, deformable kernels in both DAAB and DACB tend to gather at the equator and scatter at the polar, which conforms to the geometric distribution of pixel density in ERP images.
Besides, DAAB can also learn an overall kernel translation (the 29th DAAB), which can be regarded as a self-adaptively shift window operation.

\textbf{Pseudo ERP patches.}
In \cref{subs:dataaug}, we propose a distorted dataset DF2K-ERP to tackle over-fitting issues.
We train a standard SwinIR on diverse datasets and training schemes to study the influence of data augmentation separately.
As shown in \cref{tab:aug}, while training on ODI-SR and DF2K, distortion operations in DF2K lead to better performance.
Compared with fine-tuning on DF2K-ERP pre-trained models (two-stage), training on two datasets jointly (one-stage) shows better results.
We infer that there is a domain gap between ODI-SR and DF2K-ERP, which is caused by omitted Perspective distortion\footnote{Detailed analysis can be found in the supplementary file.}.
Moreover, the advantage of distortion modulation mechanisms in OSRT is enlarged when additional training patches are applied.
\cref{fig:overfit} proves that our data augmentation scheme overcomes the over-fitting issue and improves the reconstruction ability.

\begin{table}[t]
	\centering
	\scriptsize
	\setlength{\tabcolsep}{1.5mm}{
		\begin{tabular}{c|c|c|c|cc}
			\toprule
			Backbone & \multirow{2}{*}{Datasets} & Training & \multirow{2}{*}{Scale} & \multicolumn{2}{c}{SUN360} \\ 
			network &  & scheme &  & PSNR & SSIM  \\ \midrule
			SwinIR & ODI-SR & N/A & \multirow{5}{*}{$\times$2}  & 31.21 & 0.8852 \\ 
			SwinIR & DF2K/ODI-SR & one-stage &   & 31.26 & 0.8841 \\ 
			SwinIR & DF2K-ERP/ODI-SR & one-stage &   & 31.33 & 0.8855 \\ 
			SwinIR & DF2K-ERP/ODI-SR & two-stage &    & 31.17 & 0.8818 \\ 
			OSRT & DF2K-ERP/ODI-SR & one-stage &    & \textbf{31.52} & \textbf{0.8888} \\ \midrule
			SwinIR & ODI-SR & N/A & \multirow{5}{*}{$\times$4} & 27.39 & 0.7707 \\
			SwinIR & DF2K/ODI-SR & one-stage &  & 27.59 & 0.7768 \\ 
			SwinIR & DF2K-ERP/ODI-SR & one-stage &  & 27.71 & 0.7804 \\ 
			SwinIR & DF2K-ERP/ODI-SR & two-stage &  & 27.74 & 0.7795 \\ 
			OSRT & DF2K-ERP/ODI-SR & one-stage &   & \textbf{27.84} & \textbf{0.7835} \\ \bottomrule
		\end{tabular}
	}
	\vspace{-3mm}
	\caption{Ablation study on data augmentation. The results of ODI-SR (In the supplementary file) are in the same trend as SUN360.}
	\vspace{-5mm}
	\label{tab:aug}
\end{table}

\section{Conclusion}
In this paper, we find that the previous downsampling process in the ODISR task harms the intrinsic distribution of pixel density in ODIs, which leads to poor generalization ability in real-world scenarios.
To tackle this issue, we propose Fisheye downsampling, which mimics the real-world imaging process to preserve the realistic density distribution.
After refining the downsampling process, we design a distortion-aware Transformer (OSRT) to modulate distortions continuously and self-adaptively.
OSRT learns offsets from the distortion-related condition and rectifies distortion by feature-level warping.
Moreover, to alleviate the over-fitting problem of large networks, we propose to synthesize additional ERP training data from the plain images.
Extensive experiments have demonstrated the state-of-the-art performance of our OSRT.
\textbf{Limitation.}
This work focuses on the feature extract process in ODISR.
However, to get a better viewing experience, the process of sampling ERP images into viewing types also requires careful design.

{\small
\bibliographystyle{ieee_fullname}
\bibliography{OSRT}
}

\clearpage
\appendix

\begin{figure*}[h]
	\centering
	\vspace{-4mm}
	\includegraphics[width=1\linewidth]{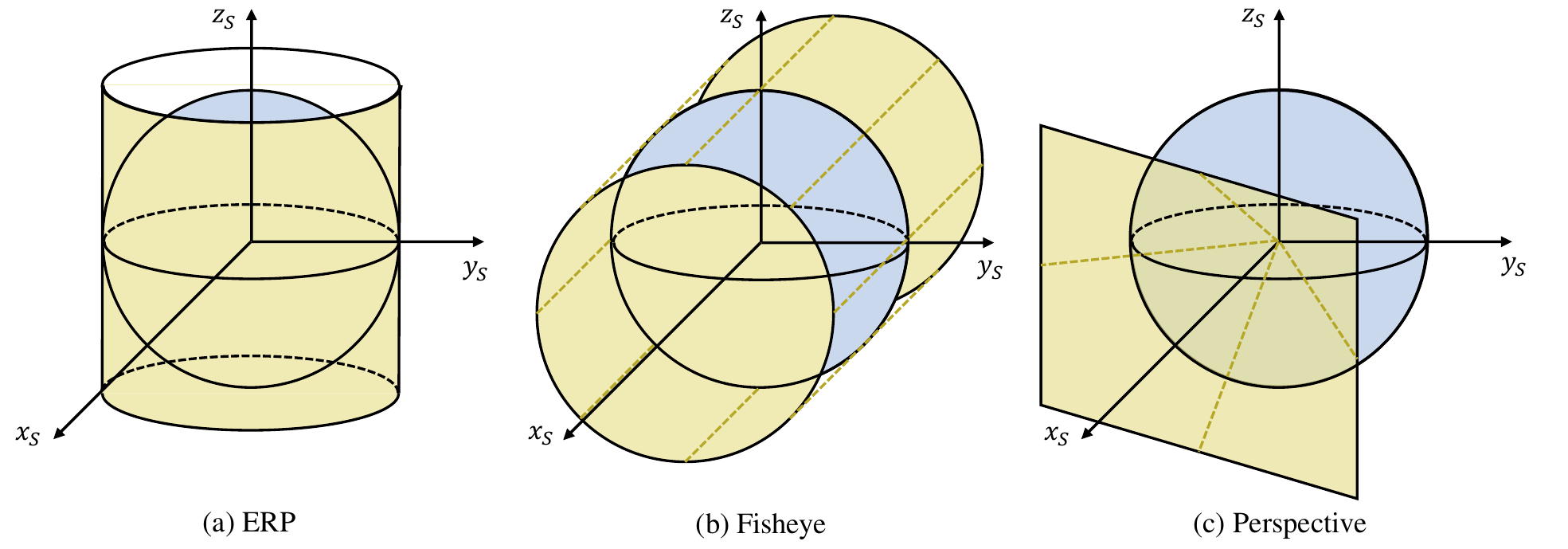}
	\vspace{-2mm}
	\caption{Geometric illustration of three projection types. Blue and yellow refer to the spherical surface and projection plane, respectively.}
	\label{fig:transform}
\end{figure*}

\noindent\textbf{\large{Appendix}} \label{appendix}

\textit{
	Due to the lack of space in the main paper, we provide more details of the proposed OSRT in the supplementary file.
	In \cref{sec:geo}, we show the transformation relationships from the uniformed sphere to various projection types (ERP, Fisheye, and Perspective) and the derivation processes of each projection type.
	More experimental details and interpretations can be found in \cref{sec:dis}.
	Then we provide additional visual comparisons and visualizations under various projection types in \cref{sec:visual}.
}

\section{Geometric Relationship}\label{sec:geo}

In this section, $x_E, y_E$ and $x_P, y_P$ refer to plane coordinates of ERP and Perspective, respectively.
For an ideal sphere, $\theta_S, \varphi_S$ are the spherical coordinates, and $x_S, y_S, z_S$ are the space coordinates.
$\rho_F,\theta_F$ and $x_F, y_F$ are polar coordinates and plane coordinates of Fisheye, respectively.

\subsection{Transformation}

\textbf{ERP.}
For ERP, the coordinate is defined as:
\begin{equation}\label{eq:erp}
	\left\{
	\begin{aligned}
		x_E = \theta_S\enspace \\
		y_E = \varphi_S.
	\end{aligned}
	\right. 
\end{equation}

\textbf{Fisheye.}
For Fisheye, the coordinate is defined as:
\begin{equation}\label{eq:fisheye_all}
	\left\{
	\begin{aligned}
		\rho_F = 2\times\arctan(\sqrt{x_S^2+y_S^2/z_S^2})/A_F \\
		\theta_F = \arctan(y_S/x_S) \\
		x_S = \rho_F \times \cos(\theta_F)\, \\
		y_S = \rho_F \times \sin(\theta_F),
	\end{aligned}
	\right.
\end{equation}
where $A_F$ is the aperture degree of Fisheye.
Specifically, when the normal vector of the Fisheye splicing plane is parallel to the z-axis, \cref{eq:fisheye_all} can be simplified as:
\begin{equation}\label{eq:fisheye_xoy}
	\left\{
	\begin{aligned}
		\rho_F = 2\times(\pi/2-\varphi_S)/A_F \\
		\theta_F = \theta_S.
	\end{aligned}
	\right.
\end{equation}
Here, we define a rotation transformation under the spherical coordinates:
\begin{equation}\label{eq:sapce_rotation}
	[x_S^*, y_S^*, z_S^*]^T = M_r\cdot[x_S, y_S, z_S]^T,
\end{equation}
where $M_r$ is the 3D rotation matrix. $[x_S, y_S, z_S]^T$ and $[x_S^*, y_S^*, z_S^*]^T$ are the original and rotated spherical coordinates, respectively.
\cref{eq:sapce_rotation} is defined to align general Fisheye to the horizontally spliced one, which is identical to add $\Delta\theta_r, \Delta\varphi_r$ on spherical polar coordinates.

\textbf{Perspective.}
The coordinates is defined as:
\begin{equation}\label{eq:pers}
	\left\{
	\begin{aligned}
		x_P = \tan(\theta_S)\; \\
		y_P = \tan(\varphi_S)/cos(\theta_S),
	\end{aligned}
	\right.
\end{equation}
where $x_P,y_P\in[-\tan(A_P/2),\tan(A_P/2)]$.
$A_P$ is the aperture degree of Perspective, which determines the field-of-view (FOV) of the given Perspective.
Note that a perspective image only represents information on a partial area of a spherical surface.

\subsection{Distortion}
As mentioned in the main paper, the distortion degree of each projection type is measured by \cite{WSPSNR}:
\begin{equation}\label{eq:k}
	\mathbf{K}(x, y)=\frac{\delta S(\theta, \varphi)}{\delta P(x, y)}=\frac{\cos (\varphi)|d \theta d \varphi|}{|d x d y|}=\frac{\cos (\varphi)}{|J(\theta, \varphi)|},
\end{equation}
where $\delta S(\cdot, \cdot)$ and $\delta P(\cdot, \cdot)$ represent the area on the spherical surface and the projection plane, respectively.
$|d i d j|$ represents a plane microunit.
$|J(\theta, \varphi)|$ is the Jacobian determinant from spherical coordinate to projection coordinate.

\textbf{ERP distortion.}
From \cref{eq:erp,eq:k}, ERP stretching ratio can be derived as:
\begin{equation}\label{eq:k_erp_supp}
	\mathbf{K}_{\operatorname{ERP}}(x_E, y_E)=\cos (\varphi_S)=\cos (y_E).
\end{equation}

\textbf{Fisheye distortion.}
In this paragraph, we denote $A_F$ as $\pi$.
$|J^*_F(\theta_S, \varphi_S)|$ can be simplified by \cref{eq:fisheye_xoy}:
\begin{equation}\label{eq:j_fisheye_xoy}
	\begin{aligned}
		&\enspace\enspace\,\, |J^*_F(\theta_S, \varphi_S)| \\
		&=
		\left|\begin{array}{ll}
			\frac{\partial(x_F)}{\partial(\theta_S)} & \frac{\partial(x_F)}{\partial(\varphi_S)} \\
			\frac{\partial(y_F)}{\partial(\theta_S)} & \frac{\partial(y_F)}{\partial(\varphi_S)}
		\end{array}\right| \\
		&=
		\left|\begin{array}{ll}
			\frac{\partial(\rho_F\cos\theta_F)}{\partial(\theta_S)} & \frac{\partial(\rho_F\cos\theta_F)}{\partial(\varphi_S)} \\
			\frac{\partial(\rho_F\sin\theta_F)}{\partial(\theta_S)} & \frac{\partial(\rho_F\sin\theta_F)}{\partial(\varphi_S)}
		\end{array}\right| \\
		&=
		\left|\begin{array}{ll}
			\frac{\partial((1-2\varphi_S/\pi)\cos\theta_S)}{\partial(\theta_S)} & 		\frac{\partial((1-2\varphi_S/\pi)\cos\theta_S)}{\partial(\varphi_S)} \\
			\frac{\partial((1-2\varphi_S/\pi)\sin\theta_S)}{\partial(\theta_S)} & 	\frac{\partial((1-2\varphi_S/\pi)\sin\theta_S)}{\partial(\varphi_S)}
		\end{array}\right| \\
		&=
		\left|\begin{array}{ll}
			-(1-2\varphi_S/\pi)\sin\theta_S & -2\cos\theta_S/\pi \\
			(1-2\varphi_S/\pi)\cos\theta_S & 	-2\sin\theta_S/\pi
		\end{array}\right| \\
		&= \frac{2}{\pi}(1-2\varphi_S/\pi)(\sin^2\theta_S + \cos^2\theta_S) \\
		&= \frac{2}{\pi}\rho_F. \\
	\end{aligned}
\end{equation}

From \cref{eq:fisheye_xoy,eq:k,eq:j_fisheye_xoy}, the stretching ratio of horizontally spliced Fisheye can be derived as:
\begin{equation}\label{eq:k_fisheye_xoy}
	\begin{aligned}
		\mathbf{K}^*_{\operatorname{Fisheye}}(x_F, y_F)&=\frac{\cos (\varphi_S)}{|J_F(\theta_S, \varphi_S)|}\\
		&=\frac{\cos (\frac{\pi}{2}(1-\rho_F))}{\frac{2}{\pi}\rho_F}.\\
	\end{aligned}
\end{equation}

Then, we can derive stretching ratio of general Fisheye from \cref{eq:k,eq:j_fisheye_xoy,eq:k_fisheye_xoy}:
\begin{equation}\label{eq:k_fisheye_supp}
	\begin{aligned}
		\mathbf{K}_{\operatorname{Fisheye}}(x_F, y_F)&=\frac{\delta S(\theta_S, \varphi_S)}{\delta P(x_F, y_F)}\\
		&=\underbrace{\frac{\delta S(\theta_S^*, \varphi_S^*)}{\delta P(x_F, y_F)}}_{\operatorname{Projection}}\cdot
		\underbrace{\frac{\delta S(\theta_S, \varphi_S)}{\delta S(\theta_S^*, \varphi_S^*)}}_{\operatorname{Rotation}}
		\\
		&=\mathbf{K}^*\cdot\frac{\cos (\varphi_S)|d \theta_S d \varphi_S|}{\cos (\varphi_S^*)|d \theta_S^* d \varphi_S^*|}\\
		&=\mathbf{K}^*\cdot\frac{\cos (\varphi_S^* + \Delta \varphi_r)}{\cos (\varphi_S^*)}\\
		&=\frac{\cos (\frac{\pi}{2}(1-\rho_F) - \Delta \varphi_r)}{\frac{2}{\pi}\rho_F},\\
	\end{aligned}
\end{equation}
where $\Delta \varphi_r$ is a constant, which is determined by the angle between the normal vector of splicing plane and z-axis.

\textbf{Perspective.}
From \cref{eq:pers,eq:k}, the Perspective stretching ratio can be derived as:
\begin{equation}\label{eq:k_pers}
	\begin{aligned}
		\mathbf{K}_{\operatorname{Perspective}}(x_P, y_P)&=\frac{\cos (\varphi_S)}{|J_P(\theta_S, \varphi_S)|}\\
		&=cos^3(\theta_S)cos^3(\varphi_S)\\
		&=(1+x^2_P+y^2_P)^{-\frac{3}{2}}.\\
	\end{aligned}
\end{equation}

\begin{table}[b]
	\centering
	\scriptsize
	\vspace{-5mm}
		\begin{tabular}{lcc}
			\toprule
			& Original & Cleaned \\ \midrule
			Num of images in ODI-SR   (training) & 1200 & 1150 \\
			Num of images in ODI-SR   (testing) & 100 & 100 \\
			Num of images in ODI-SR   (validation) & 100 & 97 \\
			Num of images in SUN360 & 100 & 100 \\
			Downsampling function & OpenCV & Pillow \\
			Downsampling target & ERP & Dual Fisheye \\
			Storage format & JPEG & PNG \\ \bottomrule
		\end{tabular}
	\vspace{-2mm}
	\caption{Differences between the original and cleaned datasets.}
	\label{tab:clean}
\end{table}

\section{Details and Discussions}\label{sec:dis}

\subsection{Data Cleaning on ODI Dataset}

\begin{figure}[]
		\small
		\footnotesize
		
		\begin{tabular}{cc}
				\hspace{-1mm}
				\includegraphics[width=0.22\textwidth]{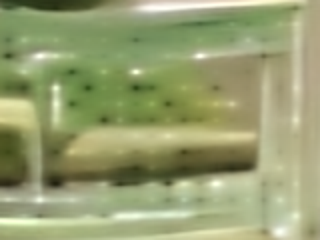} \hspace{-1mm} &
				\includegraphics[width=0.22\textwidth]{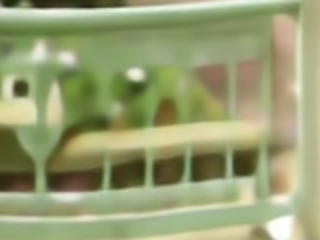} \hspace{-1mm} 
				\\
				\hspace{-1mm}
				OSRT trained on ODI-SR \hspace{-1mm} &
				OSRT trained on ODI-SR-clean  \hspace{-1mm} 
				\\
				\hspace{-1mm}
				\includegraphics[width=0.22\textwidth]{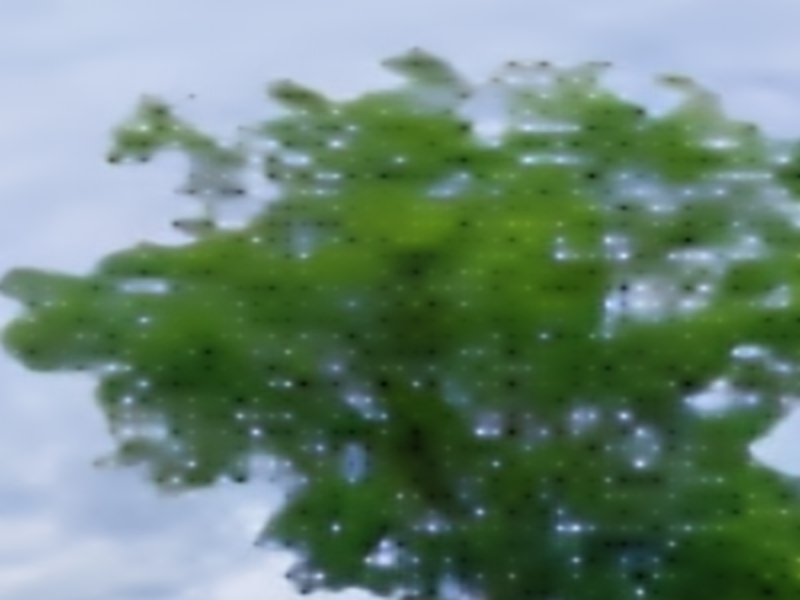} \hspace{-1mm} &
				\includegraphics[width=0.22\textwidth]{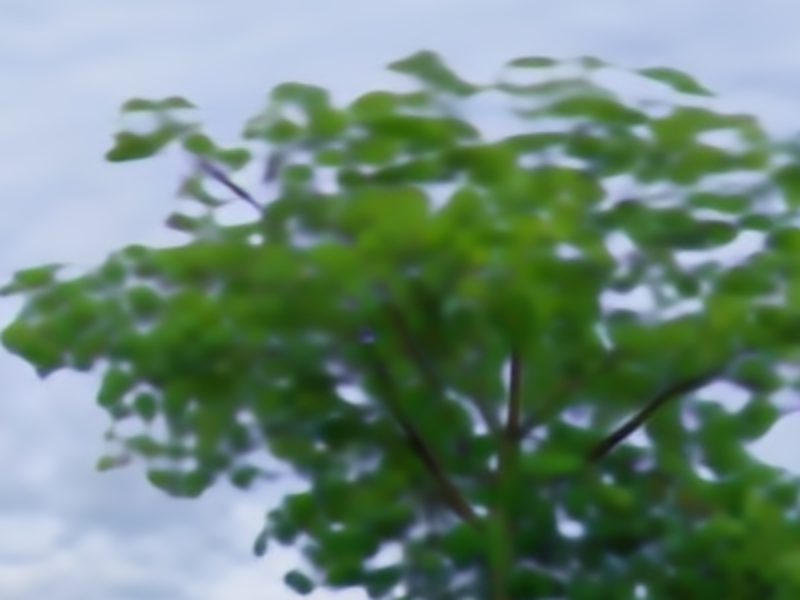} \hspace{-1mm} 
				\\
				\hspace{-1mm}
				OSRT trained on ODI-SR \hspace{-1mm} &
				OSRT trained on ODI-SR-clean  \hspace{-1mm} 
				\\	
		\end{tabular}
		\vspace{0mm}
		\caption{Visual comparisons of $\times$8 SR results trained and tested on the original and cleaned datasets.}
		\label{fig:clean}
		\vspace{-2mm}
	\end{figure}

Except for ERP downsampling, we still find other issues in both ODI-SR and SUN360 datasets.
Previous datasets are downsampled by bicubic function without anti-alias design (OpenCV-Python), which introduces mottled artifacts (\cref{fig:clean}).
Meanwhile, they are stored in the format of JPEG, which leads to missing details and JPEG-blocking artifacts.
Storing HR images in JPEG format is harmful for both training and evaluation.
To tackle these issues, we propose to apply downsampling by anti-aliased bicubic function (Pillow) and store images in a lossless format (PNG).
Moreover, there are problematic ODIs in previous datasets: \textbf{1}) transforming mistakes; \textbf{2}) virtual scenarios; \textbf{3}) extremely low qualities; \textbf{4}) plane images.
Consequently, we propose ODI-SR-clean and SUN360-clean datasets, the differences are shown in \cref{tab:clean}.
We train and test all models on cleaned datasets except the comparison under ERP downsampling (Sec. 4.3 in the main paper).

When comparing SR results under ERP downsampling, we train and test models on original datasets, which is identical to previous methods.
Thus we can directly compare the SR results of OSRT with SR results reported by previous methods, \textit{e.g.}, LAU-Net \cite{LauNet} and SphereSR \cite{SphereSR}.

\begin{table*}[t]
		\centering
		\scriptsize
		\setlength{\tabcolsep}{4mm}{
			\begin{tabular}{c|c|c|c|cc|cc}
				\toprule
				Backbone & \multirow{2}{*}{Datasets} & Training & \multirow{2}{*}{Scale} & \multicolumn{2}{c}{ODI-SR}& \multicolumn{2}{c}{SUN360} \\ 
				network &  & scheme &  & PSNR & SSIM & PSNR & SSIM  \\ \midrule
				SwinIR & ODI-SR & N/A & \multirow{5}{*}{$\times$2}  & 30.52 & 0.8819 & 31.21 & 0.8852 \\
				SwinIR & DF2K/ODI-SR & one-stage &   & 30.59 & 0.8810 & 31.26 & 0.8841 \\
				SwinIR & DF2K-ERP/ODI-SR & one-stage &   & 30.64 & 0.8821 & 31.33 & 0.8855 \\
				SwinIR & DF2K-ERP/ODI-SR & two-stage &   & 30.54 & 0.8797 & 31.17 & 0.8818 \\
				OSRT & DF2K-ERP/ODI-SR & one-stage &    & \textbf{30.77} & \textbf{0.8846} & \textbf{31.52} & \textbf{0.8888} \\ \midrule
				SwinIR & ODI-SR & N/A & \multirow{5}{*}{$\times$4} & 27.12 & 0.7663 & 27.39 & 0.7707 \\
				SwinIR & DF2K/ODI-SR & one-stage &  & 27.24 & 0.7708 & 27.59 & 0.7768 \\
				SwinIR & DF2K-ERP/ODI-SR & one-stage &  & 27.31 & 0.7735 & 27.71 & 0.7804 \\
				SwinIR & DF2K-ERP/ODI-SR & two-stage &  & 27.33 & 0.7725 & 27.74 & 0.7795 \\
				OSRT & DF2K-ERP/ODI-SR & one-stage &   & \textbf{27.41} & \textbf{0.7762} & \textbf{27.84} & \textbf{0.7835} \\ \bottomrule
			\end{tabular}
		}
		\vspace{0mm}
		\caption{Ablation study on data augmentation.}
		\vspace{-4mm}
		\label{tab:aug_supp}
	\end{table*}
	
			%
	
	\begin{table}[t]
		\centering
		\scriptsize
			\begin{tabular}{c|c|cc|cc}
				\toprule
				\multicolumn{1}{c|}{\multirow{2}{*}{Method}}             & \multirow{2}{*}{Scale} & \multicolumn{2}{c|}{ODI-SR}         & \multicolumn{2}{c}{SUN 360 Panorama} \\
				\multicolumn{1}{c|}{}&                                   & PSNR  & \multicolumn{1}{c|}{SSIM}   & PSNR & SSIM  \\ \midrule
				RCAN \cite{RCAN}                    & \multirow{2}{*}{$\times$2}    & 30.08 & 0.8723   & 30.56  & 0.8712  \\ 
				RCAN-local \cite{TLC}              &               & \textbf{30.28} & \textbf{0.8735}  & \textbf{30.80} & \textbf{0.8740}  \\ \midrule
				
				RCAN \cite{RCAN}                    &  \multirow{2}{*}{$\times$4}             & 26.85 & 0.7621   & 27.10  & 0.7660    \\ 
				RCAN-local \cite{TLC}               &              & \textbf{26.99} & \textbf{0.7622} &  \textbf{27.24} & \textbf{0.7665}   \\ \bottomrule
			\end{tabular}
		\vspace{0mm}
		\caption{Influence of test-time local converter.}
		\label{tab:tlc}
		\vspace{0mm}
	\end{table}

	\subsection{Instability of RCAN}
	
	For RCAN \cite{RCAN} trained with Fisheye downsampling, the training process is unstable and thus the performance is degraded.
	We find that the instability of RCAN is caused by incompatibility between the channel attention block (CAB) and Fisheye downsampling.
	CAB requires global statistical features, and its training stability depends on the consistent mean value distribution of each patch \cite{TLC}.
	However, when Fisheye downsampling is applied to an ERP image, the ERP image suffers from nonuniform downsampling, which directly increases the mean value diversity between patches.
	Although implementing a test-time local converter (TLC \cite{TLC}) can reduce the distribution gap between the patch and the whole image (\cref{tab:tlc}),
	it cannot reduce the distribution gap within patches.
	Consequently, while training ODISR models under Fisheye downsampling, blocks that require global statistical values are not recommended. 
	
	\subsection{Full Ablation Results of Data Augmentation}
	Due to the lack of space in the main paper, we only show partial ablation results of data augmentation strategies (Tab. 4).
	The full results are shown in \cref{tab:aug_supp}.
	Compared with fine-tuning on DF2K-ERP pre-trained models (two-stage), training on two datasets jointly (one-stage) shows better results.
	Moreover, the advantage of OSRT is enlarged when additional training patches are applied.

	\subsection{Domain Gap between Real and Pseudo ODIs}
	As mentioned in the main paper (Sec. 3.4), we synthesize pseudo ERP training data (DF2K-ERP) from the plain images to alleviate the over-fitting problem of large networks.
	Although DF2K-ERP has shown obvious benefits, there is still a domain gap between real and pseudo images.
	From \cref{eq:k_pers}, we can see that the distortion degree of Perspective is determined by the distance from the center. 
	As the projection range is determined by FOV degree, perspective images with different FOV degrees suffer inconsistent distortions.
	However, we cannot obtain the distribution of FOV degrees in real-world scenarios.
	Thus we directly assume that all pseudo perspective images have a fixed FOV degree of 90$^\circ$, which introduces a domain gap.
	While the inevitably domain gap is a limitation of DF2K-ERP, it still overcomes the over-fitting issue and improves the reconstruction ability.
	
	\begin{figure*}[b]
		\scriptsize
		\centering
		\vspace{0mm}
		\begin{tabular}{c}
			\begin{adjustbox}{valign=t}
				\begin{tabular}{c}
					\includegraphics[width=0.260\textwidth]{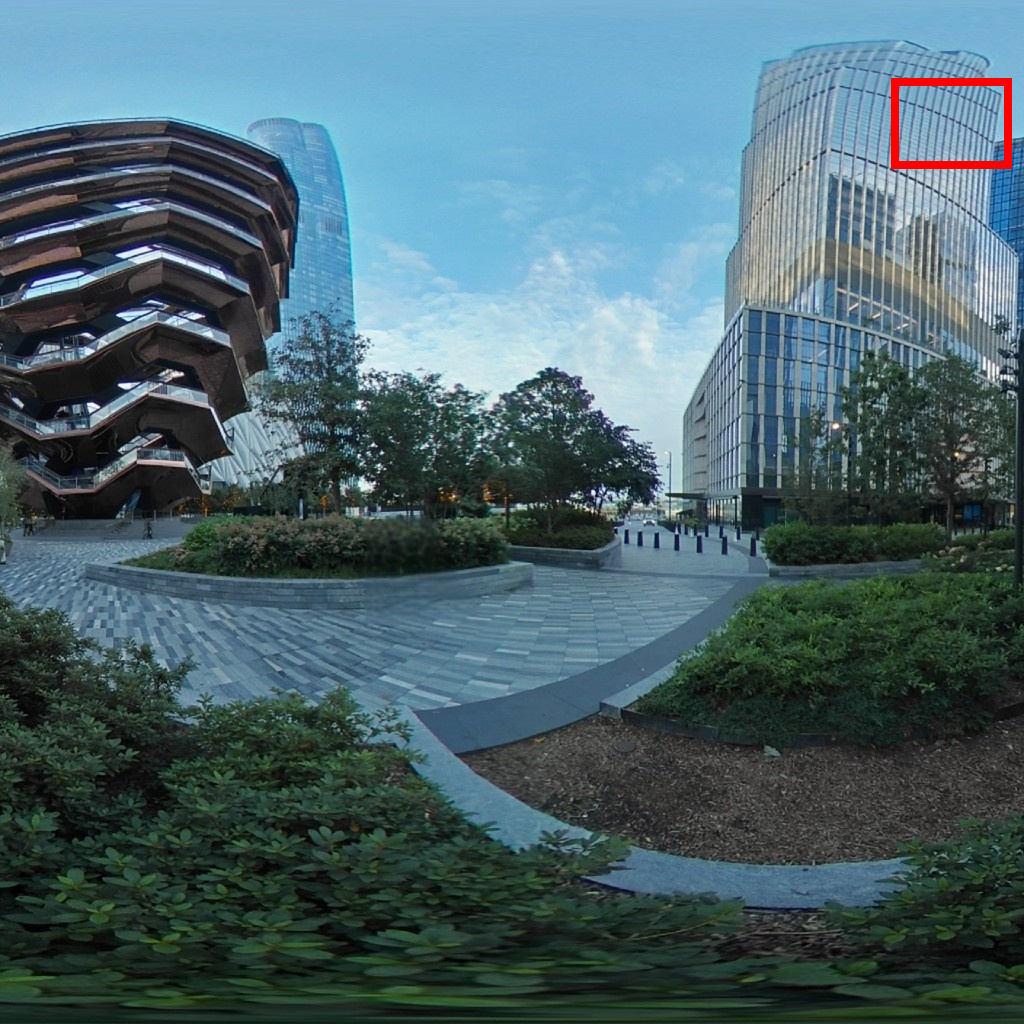}
					\\
					(a) ERP SUN360 ($\times$2): 004
				\end{tabular}
			\end{adjustbox}
			\hspace{-2mm}
			\begin{adjustbox}{valign=t}
				\begin{tabular}{cccc}
					\includegraphics[width=0.149\textwidth]{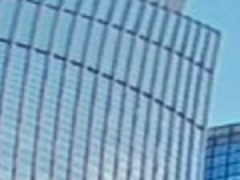} \hspace{-1mm} &
					\includegraphics[width=0.149\textwidth]{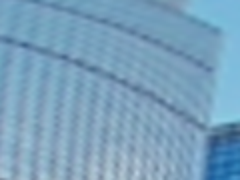} \hspace{-1mm} &
					\includegraphics[width=0.149\textwidth]{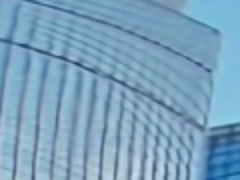} \hspace{-1mm} &
					\includegraphics[width=0.149\textwidth]{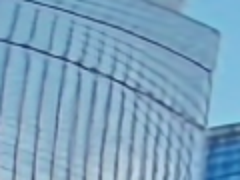} \hspace{-1mm} 
					\\
					HR \hspace{-1mm} &
					Bicubic \hspace{-1mm} &
					RCAN \cite{RCAN} \hspace{-1mm} &
					SRResNet \cite{ESRGAN} \hspace{-1mm} 
					\vspace{1.8mm}
					\\
					\includegraphics[width=0.149\textwidth]{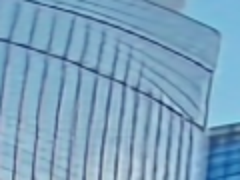} \hspace{-1mm} &
					\includegraphics[width=0.149\textwidth]{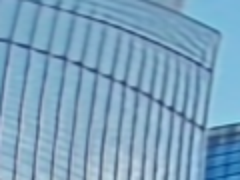} \hspace{-1mm} &
					\includegraphics[width=0.149\textwidth]{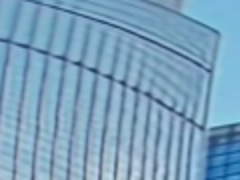} \hspace{-1mm} &
					\includegraphics[width=0.149\textwidth]{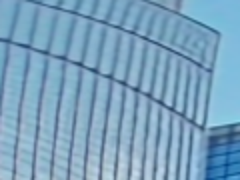} \hspace{-1mm}  
					\\ 
					EDSR \cite{EDSR} \hspace{-1mm} &
					SwinIR \cite{SwinIR} \hspace{-1mm} &
					SwinIR$^{\dagger}$ \cite{SwinIR} \hspace{-1mm} &
					OSRT$^{\dagger}$ \hspace{-1mm} 
				\end{tabular}
			\end{adjustbox}
			
			\vspace{3mm}
			\\ 
			\begin{adjustbox}{valign=t}
				\begin{tabular}{c}
					\includegraphics[width=0.260\textwidth]{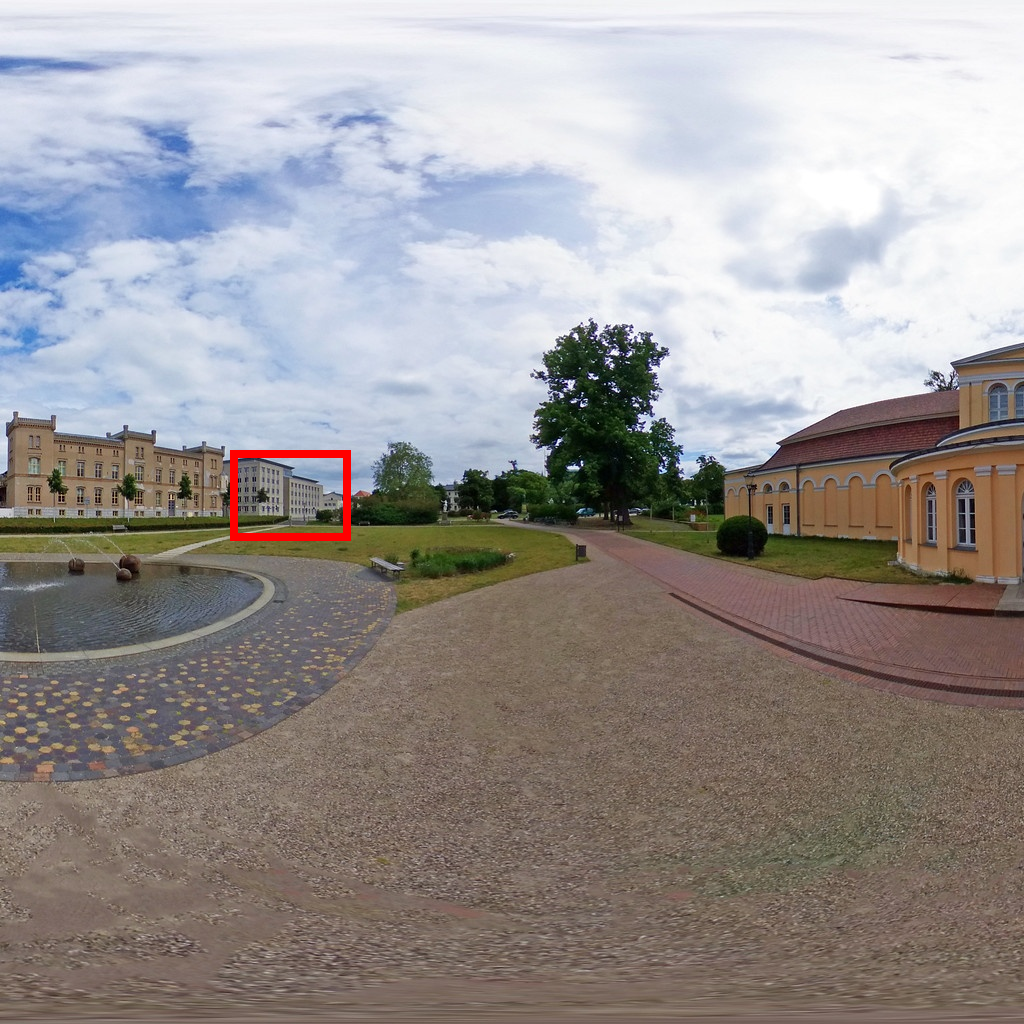}
					\\
					(b) ERP ODI-SR ($\times$4): 049
				\end{tabular}
			\end{adjustbox}
			\hspace{-2mm}
			\begin{adjustbox}{valign=t}
				\begin{tabular}{cccc}
					\includegraphics[width=0.149\textwidth]{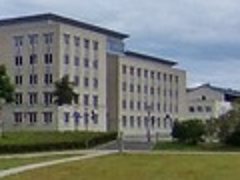} \hspace{-1mm} &
					\includegraphics[width=0.149\textwidth]{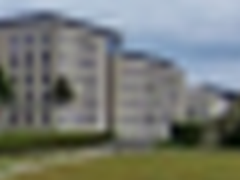} \hspace{-1mm} &
					\includegraphics[width=0.149\textwidth]{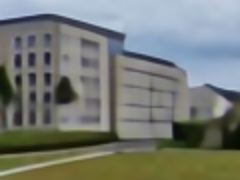} \hspace{-1mm} &
					\includegraphics[width=0.149\textwidth]{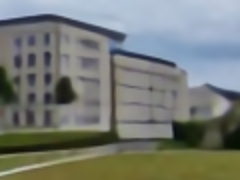} \hspace{-1mm} 
					\\
					HR \hspace{-1mm} &
					Bicubic \hspace{-1mm} &
					RCAN \cite{RCAN} \hspace{-1mm} &
					SRResNet \cite{ESRGAN} \hspace{-1mm} 
					\vspace{1.8mm}
					\\
					\includegraphics[width=0.149\textwidth]{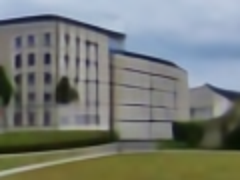} \hspace{-1mm} &
					\includegraphics[width=0.149\textwidth]{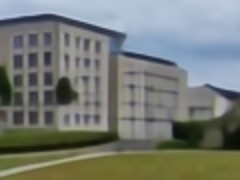} \hspace{-1mm} &
					\includegraphics[width=0.149\textwidth]{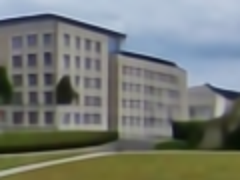} \hspace{-1mm} &
					\includegraphics[width=0.149\textwidth]{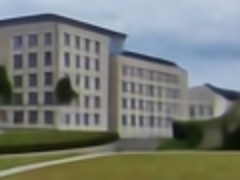} \hspace{-1mm}  
					\\ 
					EDSR \cite{EDSR} \hspace{-1mm} &
					SwinIR \cite{SwinIR} \hspace{-1mm} &
					SwinIR$^{\dagger}$ \cite{SwinIR} \hspace{-1mm} &
					OSRT$^{\dagger}$ \hspace{-1mm} 
				\end{tabular}
			\end{adjustbox}
			
			\vspace{3mm}
			\\ 
			\begin{adjustbox}{valign=t}
				\begin{tabular}{c}
					\includegraphics[width=0.46\columnwidth]{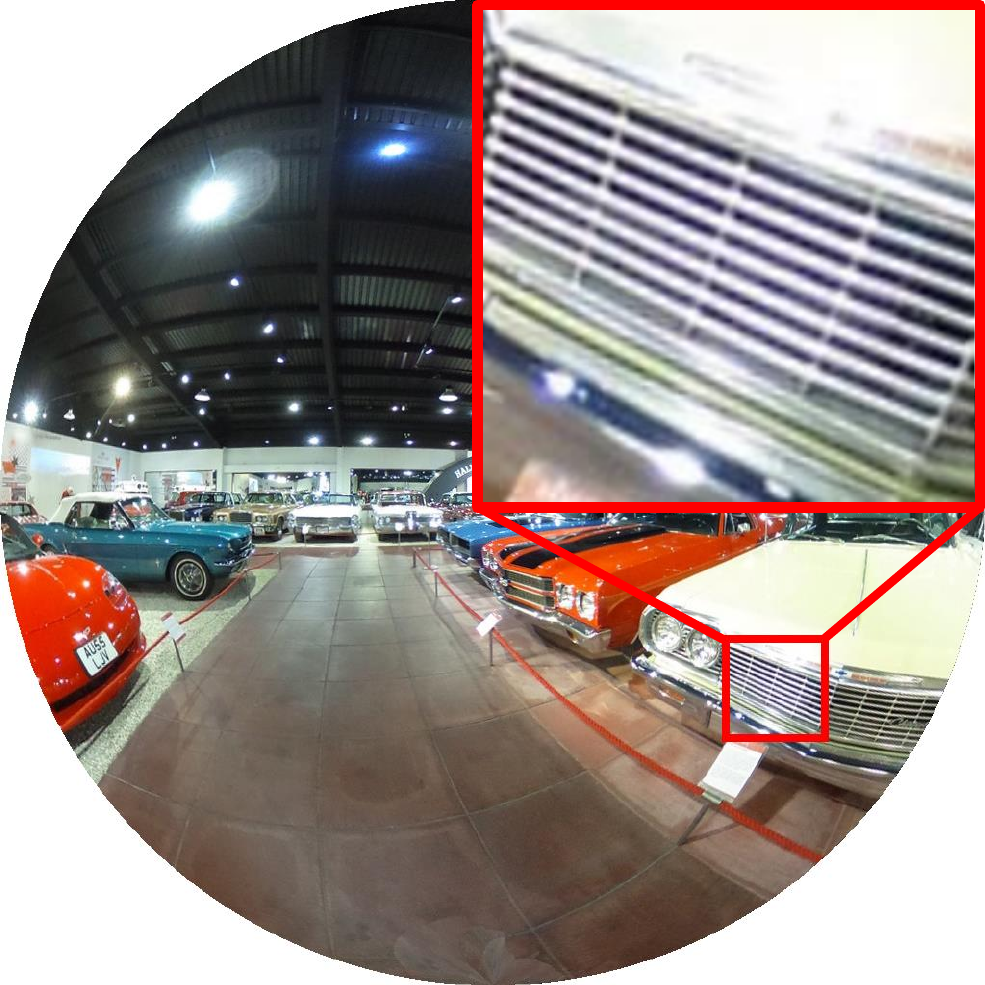}
					\\
					(c) ODI-SR ($\times$4): 003
					\\
					Fisheye (Vertical, Right)
				\end{tabular}
			\end{adjustbox}
			\hspace{-3mm}
			\begin{adjustbox}{valign=t}
				\begin{tabular}{cc}
					\includegraphics[width=0.23\columnwidth]{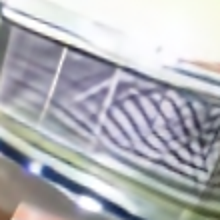} \hspace{-3mm} &
					\includegraphics[width=0.23\columnwidth]{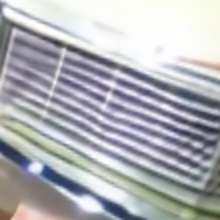} \hspace{-1mm} 
					\\
					EDSR \cite{EDSR} \hspace{-3mm} &
					SwinIR \cite{SwinIR} \hspace{-1mm} 
					\\
					\includegraphics[width=0.23\columnwidth]{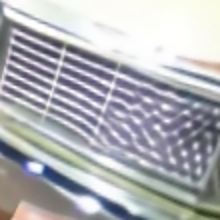} \hspace{-3mm} &
					\includegraphics[width=0.23\columnwidth]{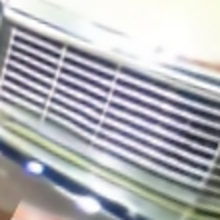} \hspace{-1mm} 
					
					\\ 
					SwinIR$^{\dagger}$ \cite{SwinIR} \hspace{-3mm} &
					OSRT$^{\dagger}$ \hspace{-1mm} 
					
				\end{tabular}
			\end{adjustbox}
			
			\begin{adjustbox}{valign=t}
				\begin{tabular}{c}
					\includegraphics[width=0.46\columnwidth]{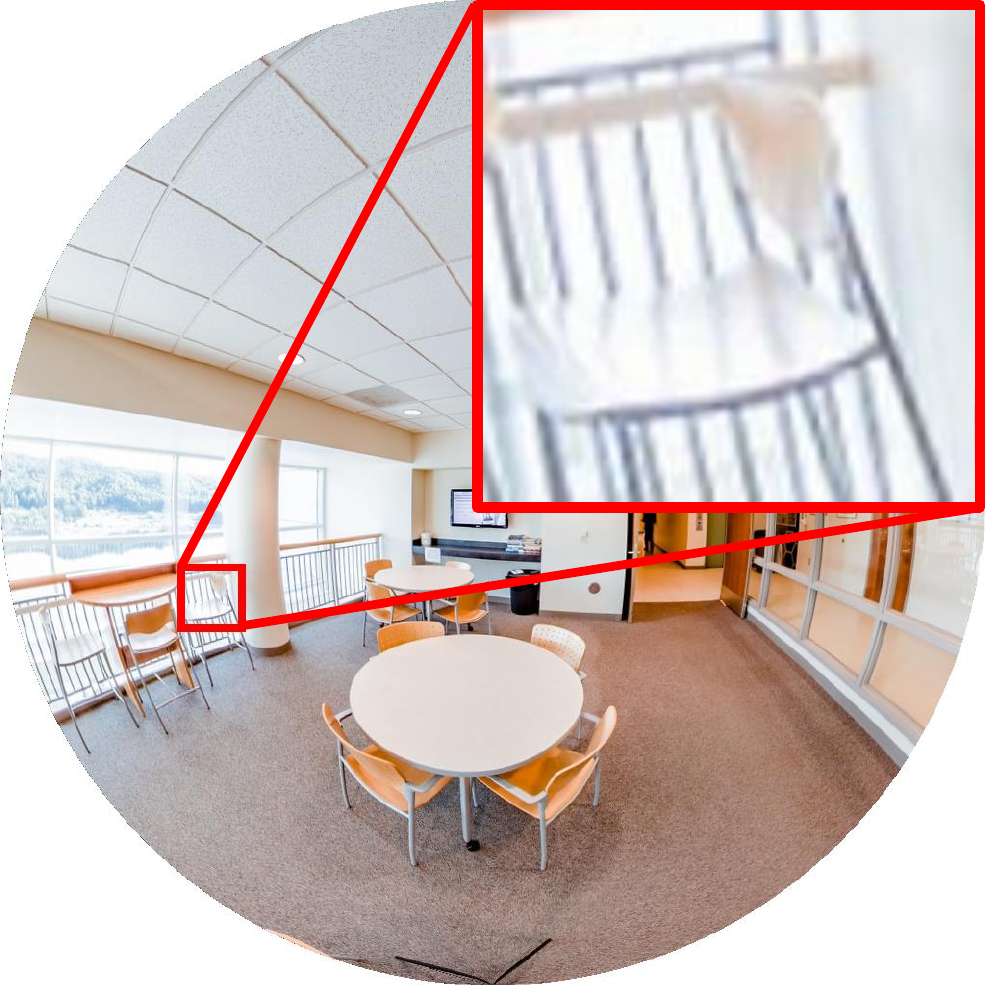}
					\\
					(d) SUN360 ($\times$4): 047
					\\
					Fisheye (Vertical, Left)
				\end{tabular}
			\end{adjustbox}
			\hspace{-3mm}
			\begin{adjustbox}{valign=t}
				\begin{tabular}{cc}
					\includegraphics[width=0.23\columnwidth]{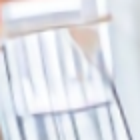} \hspace{-3mm} &
					\includegraphics[width=0.23\columnwidth]{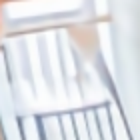} \hspace{-1mm} 
					\\
					EDSR \cite{EDSR} \hspace{-3mm} &
					SwinIR \cite{SwinIR} \hspace{-1mm} 
					\\
					\includegraphics[width=0.23\columnwidth]{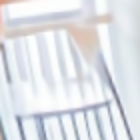} \hspace{-3mm} &
					\includegraphics[width=0.23\columnwidth]{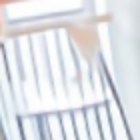} \hspace{-1mm} 
					
					\\ 
					SwinIR$^{\dagger}$ \cite{SwinIR} \hspace{-3mm} &
					OSRT$^{\dagger}$ \hspace{-1mm} 
					
				\end{tabular}
			\end{adjustbox}
			
			\vspace{3mm}
			\\
			\begin{adjustbox}{valign=t}
				\begin{tabular}{c}
					\includegraphics[width=0.46\columnwidth]{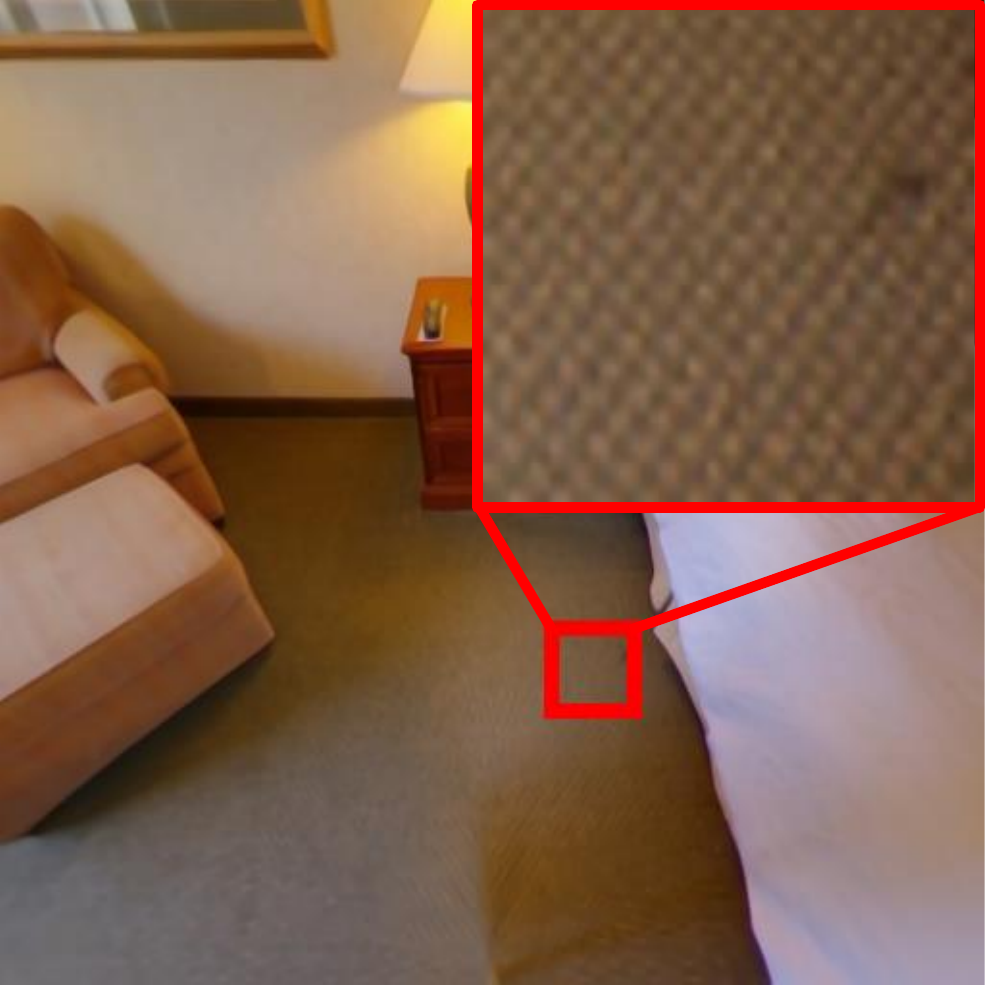}
					\\
					(e) SUN360 ($\times$4): 096
					\\
					Perspective ($\varphi$: $-45^{\circ}$; FOV: $90^{\circ}$)
				\end{tabular}
			\end{adjustbox}
			\hspace{-3mm}
			\begin{adjustbox}{valign=t}
				\begin{tabular}{cc}
					\includegraphics[width=0.23\columnwidth]{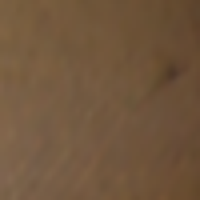} \hspace{-3mm} &
					\includegraphics[width=0.23\columnwidth]{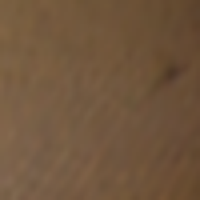} \hspace{-1mm} 
					\\
					EDSR \cite{EDSR} \hspace{-3mm} &
					SwinIR \cite{SwinIR} \hspace{-1mm} 
					\\
					\includegraphics[width=0.23\columnwidth]{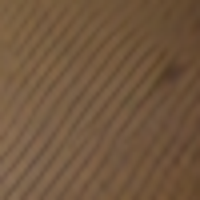} \hspace{-3mm} &
					\includegraphics[width=0.23\columnwidth]{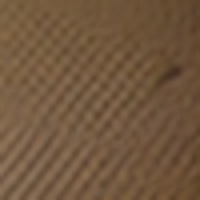} \hspace{-1mm} 
					
					\\ 
					SwinIR$^{\dagger}$ \cite{SwinIR} \hspace{-3mm} &
					OSRT$^{\dagger}$ \hspace{-1mm} 
					
				\end{tabular}
			\end{adjustbox}
			\begin{adjustbox}{valign=t}
				\begin{tabular}{c}
					\includegraphics[width=0.46\columnwidth]{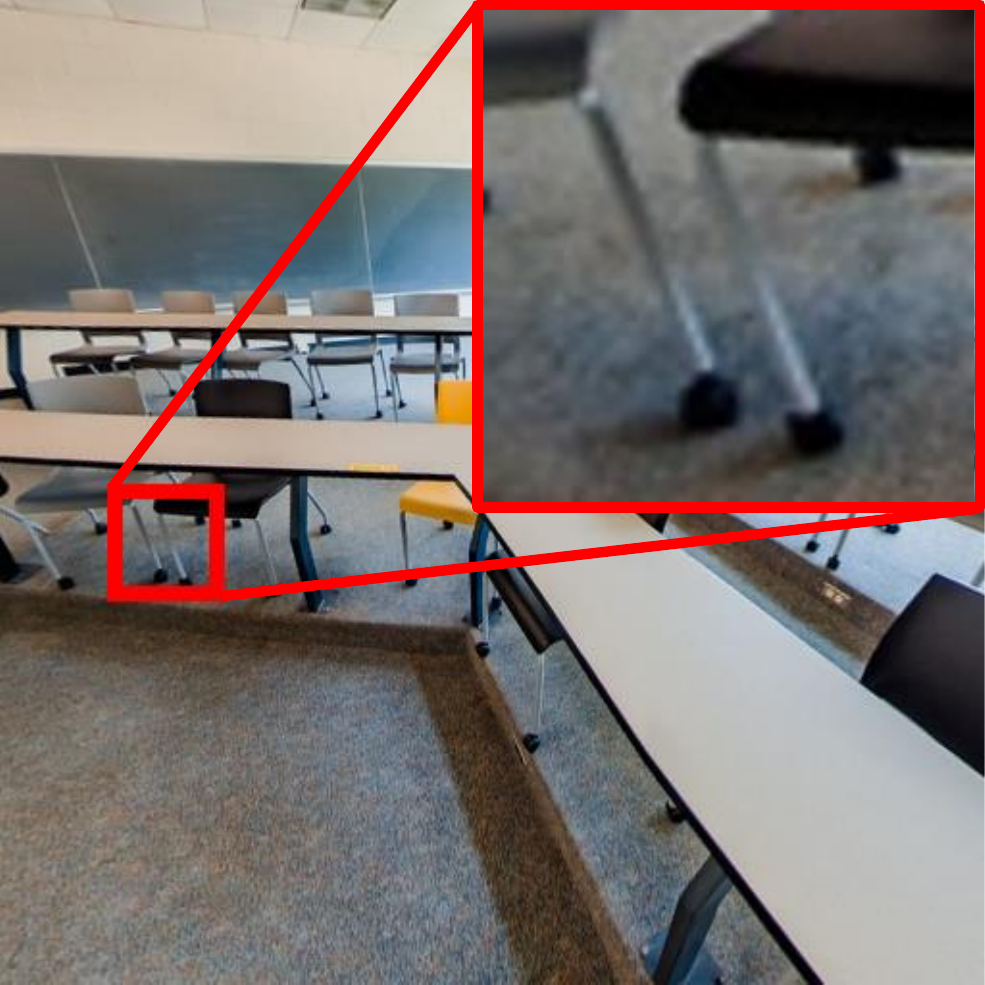}
					\\
					(f) SUN360 ($\times$4): 032
					\\
					Perspective ($\varphi$: $-30^{\circ}$; FOV: $90^{\circ}$)
				\end{tabular}
			\end{adjustbox}
			\hspace{-3mm}
			\begin{adjustbox}{valign=t}
				\begin{tabular}{cc}
					\includegraphics[width=0.23\columnwidth]{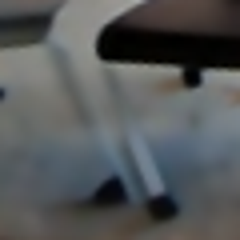} \hspace{-3mm} &
					\includegraphics[width=0.23\columnwidth]{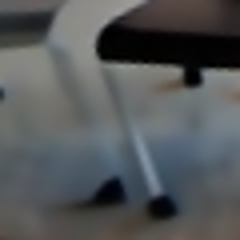} \hspace{-1mm} 
					\\
					EDSR \cite{EDSR} \hspace{-3mm} &
					SwinIR \cite{SwinIR} \hspace{-1mm} 
					\\
					\includegraphics[width=0.23\columnwidth]{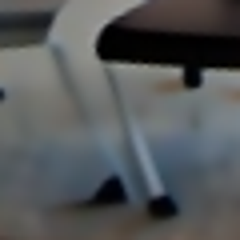} \hspace{-3mm} &
					\includegraphics[width=0.23\columnwidth]{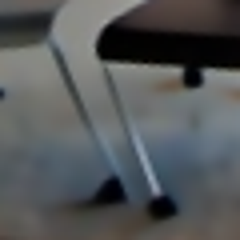} \hspace{-1mm} 
					
					\\ 
					SwinIR$^{\dagger}$ \cite{SwinIR} \hspace{-3mm} &
					OSRT$^{\dagger}$ \hspace{-1mm} 
					
				\end{tabular}
			\end{adjustbox}

		\end{tabular}
		\vspace{0mm}
		\caption{Visual comparisons of SR results under Fisheye downsampling. $\dagger$ denotes applying DF2K-ERP as augmented dataset.}
		\label{fig:supp_x4_visual}
	\end{figure*}
	
	\section{Visualization}\label{sec:visual}
	As mentioned in the main paper (Sec. 3.2), ERP downsampling leads to unrealistic ODIs.
	Thus we only show visualizations based on Fisheye downsampling in this section.
	
	\textbf{Additional qualitative comparison.}
	We provide additional visual comparisons with other methods on the ODI-SR-clean testing dataset and SUN360-clean dataset in \cref{fig:supp_x4_visual}.
	Reconstructed ERP images are compared under ERP, Fisheye, and Perspective.
	As shown in \cref{fig:supp_x4_visual} (d) and (f), we can see that OSRT can reconstruct sharp and accurate boundaries.
	Besides, from \cref{fig:supp_x4_visual} (a) and (c), we conclude that OSRT is skilled at reconstructing rigid textures.
	
\begin{figure*}[p]
		\centering
		\resizebox{\textwidth}{!}{
			\includegraphics[]{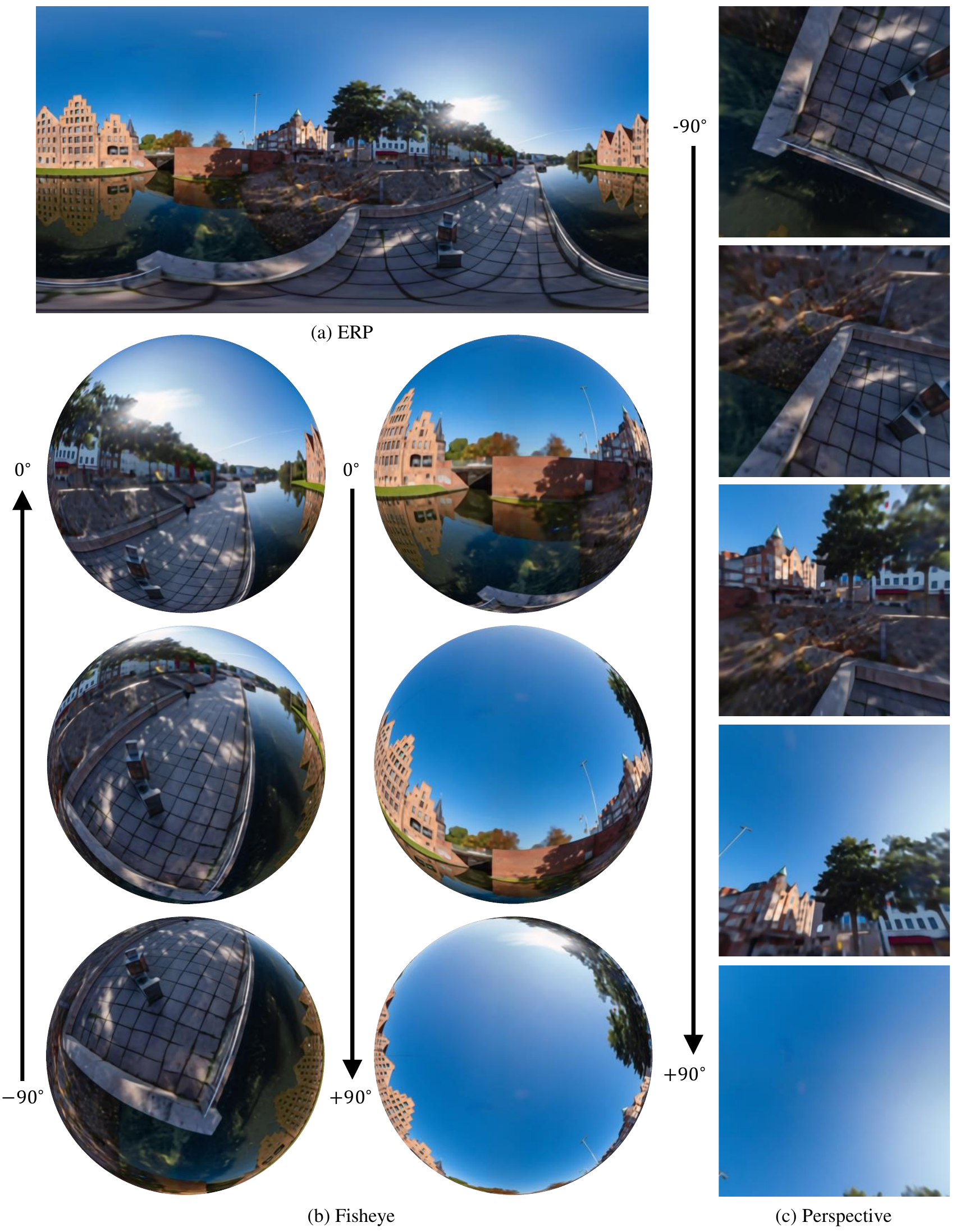}
		}
		\vspace{-7mm}
		\caption{Visualization of $\times$8 SR results (SUN360-062).}
		\label{fig:visual_full_x8}
	\end{figure*}

\begin{figure*}[p]
		\centering
		\resizebox{\textwidth}{!}{
			\includegraphics[]{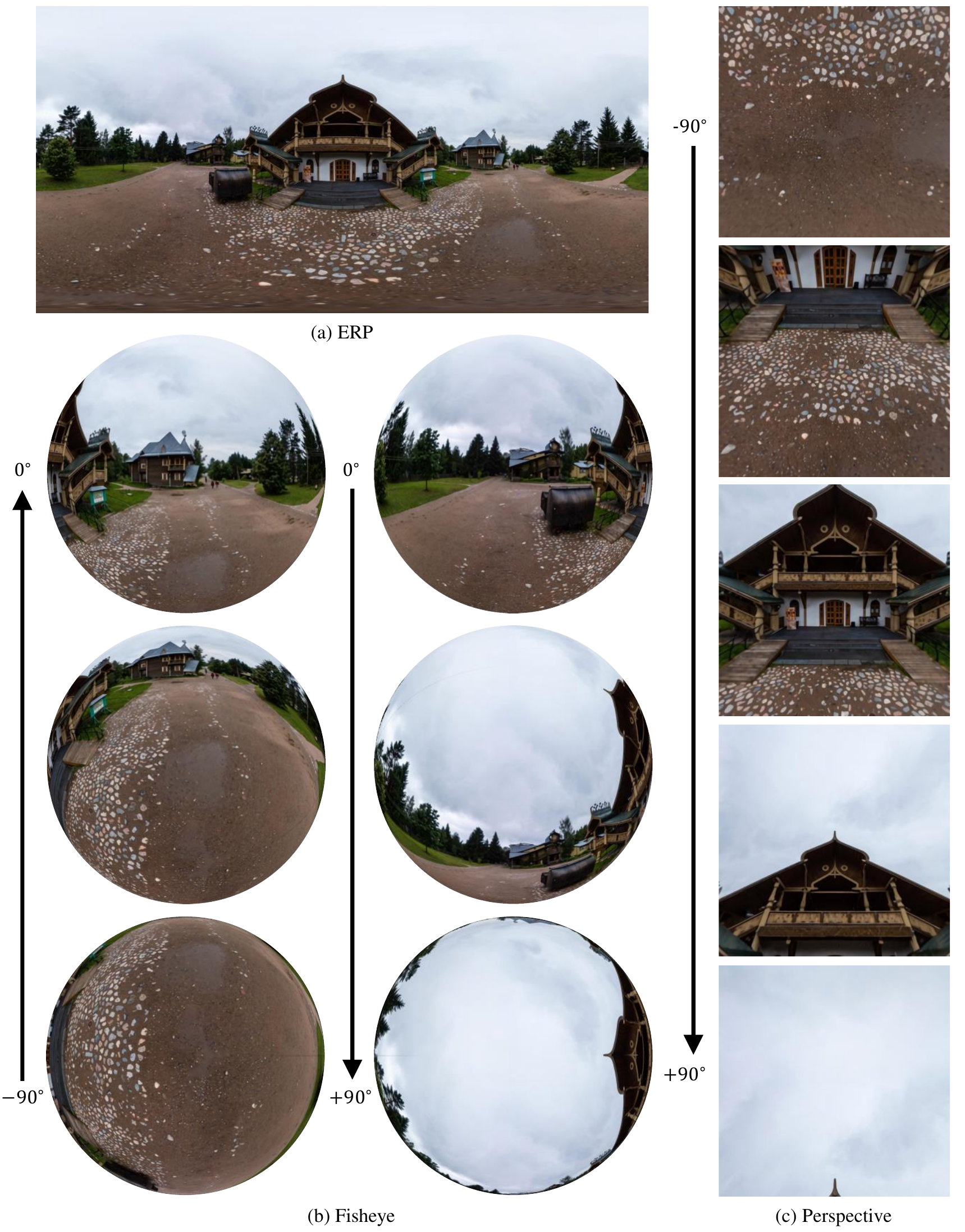}
		}
		\vspace{-7mm}
		\caption{Visualization of $\times$4 SR results (ODI-SR-066).}
		\label{fig:visual_full_x4}
	\end{figure*}

\begin{figure*}[p]
		\centering
		\resizebox{\textwidth}{!}{
			\includegraphics[]{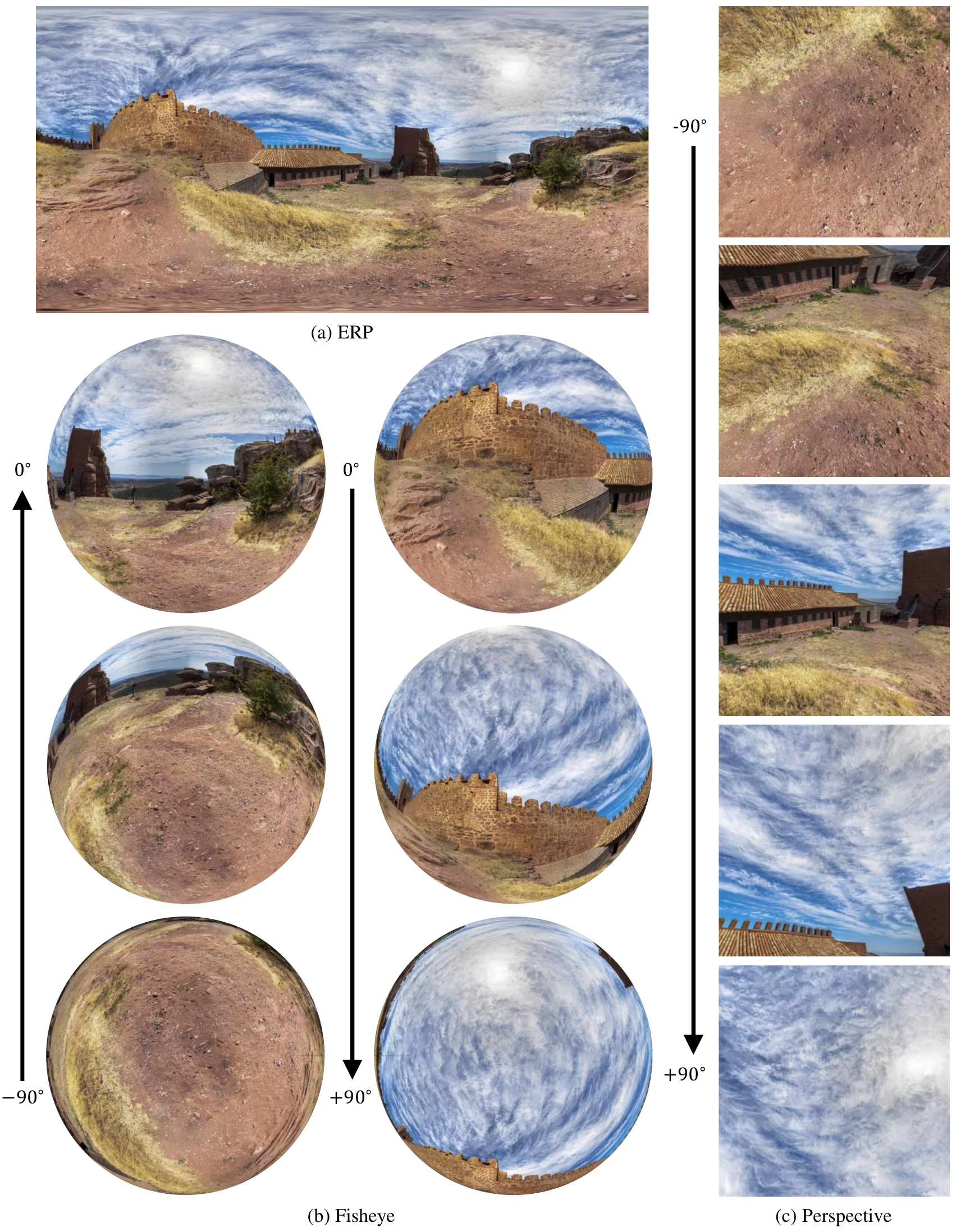}
		}
		\vspace{-7mm}
		\caption{Visualization of $\times$2 SR results (SUN360-007).}
		\label{fig:visual_full_x2}
	\end{figure*}

	\textbf{Additional visualization of OSRT.}
	To show the overall quality of OSRT reconstructed images, we project these ERP images to arbitrary projection types.
	\cref{fig:visual_full_x8,fig:visual_full_x4,fig:visual_full_x2} depict visualizations of $\times$2, $\times$4 and $\times$8 SR results, respectively.
	Under all projection types, OSRT can reconstruct details with high fidelity (buildings in \cref{fig:visual_full_x8}, tiles in \cref{fig:visual_full_x4}, and grasses in \cref{fig:visual_full_x2}).

\end{document}